\documentclass[a4paper,11pt]{article}
\pdfoutput=1 

\usepackage{jheppub} 
\usepackage{ytableau}
\ytableausetup{boxsize=0.9em}
\usepackage[T1]{fontenc} 
\usepackage{latexsym}

\usepackage{extarrows}

\usepackage{xcolor}

\usepackage{graphicx} 

\usepackage{booktabs} 
\usepackage{array} 
\usepackage{paralist} 
\usepackage{verbatim} 
\usepackage{subfig} 

\usepackage{fancyhdr} 
\usepackage{sectsty}
\allsectionsfont{\sffamily\mdseries\upshape} 

\usepackage[nottoc,notlof,notlot]{tocbibind} 
\usepackage[titles,subfigure]{tocloft} 

\newcommand{\tabincell}[2]{\begin{tabular}{@{}#1@{}}#2\end{tabular}}

\usepackage{amsfonts}

\usepackage{amsmath}
\usepackage{amssymb}
\DeclareMathOperator{\Tr}{Tr}

\preprint{USTC-ICTS/PCFT-25-45}
\title{\boldmath A graphical representation of gluonic operators}

\author[a]{Qingjun Jin,}
\author[b]{Ke Ren,}
\author[c,d,e]{Gang Yang,}
\author[f]{Rui Yu}

\affiliation[a]{Graduate School of China Academy of Engineering Physics,
No. 10 Xibeiwang East Road, Haidian District, Beijing, 100193, China}
\affiliation[b]{School of Physics and Astronomy, Sun Yat-Sen University, Zhuhai 519082, China}
\affiliation[c]{Institute of Theoretical Physics,
Chinese Academy of Sciences, Beijing 100190, China}
\affiliation[d]{School of Fundamental Physics and Mathematical Sciences, Hangzhou Institute for Advanced Study, UCAS, Hangzhou 310024, China}
\affiliation[e]{Peng Huanwu Center for Fundamental Theory, Hefei, Anhui 230026, China}
\affiliation[f]{School of Physical Science and Technology, Inner Mongolia University, Hohhot 010021, China}

\emailAdd{qjin@gscaep.ac.cn}
\emailAdd{renk9@mail.sysu.edu.cn}
\emailAdd{yangg@itp.ac.cn}
\emailAdd{yurui@imu.edu.cn}

\abstract{Composite local operators are central to effective field theories (EFTs), as they define interaction vertices in effective Lagrangians and play a fundamental role in investigating the structure of quantum field theories. The contribution of high-dimensional operators in the Standard Model Effective Field Theory (SMEFT) grows increasingly important as experimental precision improves at the Large Hadron Collider (LHC) and in future colliders. However, the number of operators increases very rapidly with dimension, making it extremely challenging to identify their complete set.
In our previous work \cite{Jin:2020pwh}, we proposed a systematic method for generating gluonic operators using primitive operators. In this paper, we introduce a graphical representation of gluonic operators and, based on this representation, present a method to systematically construct primitive operators. Using this method, we derive primitive operators corresponding to gluonic operators of length 2 to length 7 in $D$-dimensions.}

\begin{document} 
\maketitle
\flushbottom

\section{Introduction}

Effective Field Theory (EFT) is a powerful framework for studying the low-energy behavior of physical systems by decoupling disparate energy scales \cite{Fermi:1934hr, Huggett:1995vya, vanKolck:1999mw, Weinberg:1980wa}. Its core principle lies in the systematic separation of dynamics across different scales: when a system contains particles with widely separated masses (e.g., light particles with $M\ll \Lambda$ and heavy particles with $M\sim \Lambda$, where $\Lambda$ is the EFT cutoff scale), the high-energy degrees of freedom (heavy particles) can be integrated out at energies far below $\Lambda$. Their effects are then encoded in local operators constructed solely from the light degrees of freedom within an effective Lagrangian. This scale-separation approach not only simplifies calculations but also provides a systematic framework to parameterize new physics effects in a model-independent manner.

A prominent example is the Standard Model Effective Field Theory (SMEFT), which extends the Standard Model (SM) Lagrangian with higher-dimensional operators invariant under the SM gauge symmetry $SU(3)_C\times SU(2)_L\times U(1)_Y$:
\begin{equation}
\mathcal{L}_{\text{SMEFT}}=\mathcal{L}_{\text{SM}}+\sum_{\triangle\ge5}\sum_i \frac{C_i^{(\triangle)}}{\Lambda^{\triangle-4}}\mathcal{O}_i^{(\triangle)}\ ,
\end{equation}
where $\triangle$ denotes the mass dimension of the operators  and the Wilson coefficients $C_i^{(\triangle)}$ encode physics above the cutoff scale $\Lambda$.
SMEFT serves as a universal bridge between high- and low-energy experiments. For example, collider data can be correlated with low-energy precision tests (e.g., neutron decay or flavor-changing processes) to globally constrain new physics parameter spaces (see e.g. \cite{Gupta:2012mi, Dawson:2013bba, LHCHiggsCrossSectionWorkingGroup:2016ypw, D0:2012kms, Boggia:2017hyq}).
Complete and non-redundant operator bases have been established for dimensions 5-9 \cite{Weinberg:1979sa, Abbott:1980zj, Buchmuller:1985jz, Grzadkowski:2010es, Lehman:2014jma, Liao:2016hru, Li:2020gnx, Murphy:2020rsh, Li:2020xlh, Liao:2020jmn}.

Traditional algebraic methods for constructing high-dimensional gluon operators face challenges such as difficulty in eliminating redundancy and complications in dimensional extension, necessitating the development of new and more efficient operator construction approaches.
In this work, we will focus on pure gluon operators which are Lorentz invariant in $D$-dimensions. 
These operators describe gluon self-interactions in SMEFT. When combined with the Higgs field, they can also be regarded as the interactions of Higgs with gluons \cite{Wilczek:1977zn, Shifman:1979eb, Dawson:1990zj, Djouadi:1991tka, Kniehl:1995tn, Chetyrkin:1997sg} after integrating out the heavy top quark field:\footnote{The term "Higgs EFT" also refers to a complete distinct EFT, (see e.g. \cite{Feruglio:1992wf}) which  is similar to SMEFT, but the physical Higgs and Goldstone bosons are not treated as a doublet. It shall not be confused with the Higgs EFT here, which describes the effective interaction between Higgs and gluons after integrating out heavy quarks.}
\begin{equation}
\mathcal{L}_{\text{HEFT}}=\frac{1}{M_t}\hat{C}_{5,1}H\Tr(F^2)+\frac{1}{M_t^3}\hat{C}_{7,1}H\Tr(F^3)+\frac{1}{M_t^3}\hat{C}_{7,2}H\partial^2\Tr(F^2)+\cdots
\end{equation}

Our motivations are twofold. First, the SMEFT operators are usually constructed in spacetime dimension 4, and the evanescent operators, which vanish in 4-dimension, are neglected. 
Evanescent operators play a crucial role in the matching conditions for Wilson coefficients within the EFT framework, and the calculation of anomalous dimensions for physical (4-dimensional) operators, specifically at two-loop level and above \cite{Buras:1989xd, Dugan:1990df, Herrlich:1994kh,Buras:1998raa, Jin:2022ivc,Jin:2022qjc, Jin:2023cce, Jin:2023fbz}.
We construct the complete set of gluon operators which preserves $D$-dimensional Lorentz invariance, including both physical and evanescent operators, thus facilitating the study of  multi-loop matching and renormalization for gluon operators.\footnote{There are also operators which only preserve 4-dimensional Lorentz invariance, which involves 4-d Levi-Civita tensor $\epsilon_{\mu\nu\rho\sigma}$. We will leave the discussion of these operators to future work.}

Second, we hope to show that the construction of operators can be enhanced by using the structures and symmetries of high dimensional operators. Different gluon operators are related by permutation of fields, and can be classified by the representations of $S_n$ group. Operators with different mass dimensions are also related: the infinite tower of operators with a certain length can be generated by a finite number of "primitive operators".

It can be a non-trivial task to find the independent gluon operators, since they satisfy relations like Bianchi identies and equation of motion (EOM). The minimal form factors have a one-to-one correspondence with the operators, and they automatically satisfies the Bianchi identies and EOM. Therefore the basis of operators can be determined by examining the minimal form factors of candidate operators. 

We will further split each operator into a color factor and a kinematic part, and the latter will be called the kinematic operator. The number of kinematic operators is much smaller than the number of full operators. 
After the basis of kinematic operators has been found, the basis of color-dressed operators can be generated by dressing appropriate color factors to the basis of kinematic operators.

The number of operator, and consequently the degree of difficulty to find the operator basis, also grow drastically as mass dimension increases. The method of primitive operators \cite{Jin:2020pwh} provides a way to construct the operators with higher mass dimensions from operators with lower mass dimensions.
Primitive operators are a set of independent operators without pairs of $D_{\mu}$-$D^{\mu}$  insertions ($D$-pairs).
For a given length, there are only a finite number of primitive operators, and the complete set of operators can be constructed by adding $D$-pairs to the primitive operators.
For example, at length-2 there is a single primitive operator $\Tr(F_{\mu\nu}F^{\mu\nu})$, and 
all length-2 operators can be obtained by adding pairs of  to this primitive operator:
$D_{\rho_1\cdots\rho_k}F_{\mu\nu}D^{\rho_1\cdots\rho_k}F_{\mu\nu}$. 
The primitive operator method was also used in \cite{Jin:2022ivc} to construct the basis of evanescent gluon operators.

The length-2 and length-3 primitive operators are given in \cite{Jin:2020pwh}. In order to construct the higher length primitive operators systematically, we propose a diagrammatic representation of kinematic operators called operator diagrams. Only a few operator diagrams correspond to primitive operators, and these primitive diagrams can be easily found by examining special subdiagrams called skeleton diagrams.
Furthermore, the operator diagram representation also makes the symmetries of kinematic operators manifest, which is very helpful when we study the $S_n$ representations of kinematic operators.



This paper is organized as follows. In Section \ref{section: ko}, we start by introducing some definitions and conventions of kinematic operators. 
In Section \ref{section:diagram}, we introduce the operator diagram method. In Section \ref{subsection:skeleton}, we propose an algorithm to constructed the primitive operators. In Section \ref{section:primitive-basis}, we construct the kinematic operators with length 2 to 7 and refine them using the symmetry of the operator diagrams.  In Section \ref{section:discussion} we conclude by discussing the construction of color dressed operator basis, and some possible future directions.

\section{The kinematic part of gluon operators}
\label{section: ko}

We consider pure gluon operators which are invariant under gauge and Lorentz transformations. These operators can be constructed using convariant derivatives $D_{\mu}$ and field strength tensors $F_{\mu\nu}^a$, in which all Lorentz indices $\mu$ are contracted among $D$ and $F$, and all color indices $a_i$ must be contracted to some color factors like $\Tr(T^{a_1}\cdots T^{a_n})$.

In this section, we start by discussing some properties of gluon operators in Section \ref{subsection:gluon-operator}, then we introduce kinematic operators in Section \ref{subsection:kinematic-operator}. In  Section \ref{subsection:constraint} we discuss the physical constraints which can be used to remove redundant kinematic operators. Last we introduce the primitive operators in Section \ref{subsection:primitive}, which is the key idea in the contruction of operator basis.

\subsection{The gluon operators}
\label{subsection:gluon-operator}
Let us start by listing some properties of gluon operators, and introduce some conventions.

\subsubsection*{Abbreviation of Lorentz indices}
The gluon operators usually contains many Lorentz indices, and we would like to simplify the expressions by replacing Lorentz indices by numbers, and ignoring the difference between upper and lower indices. For example, $F^{\mu\nu } D_{\mu}F_{\rho\sigma} D_{\nu}F^{\rho\sigma }$ will be abbreviated to $ F_{12}D_1F_{34}D_2F_{34}$.

\subsubsection*{Mass dimensions}
The mass dimension of an operator is given by $\triangle=N_D+2N_F$, in which $N_D$ and $N_F$ are the number of $D$ and $F$ in the operator, respectively.
$\triangle$ is always even, because the covariant derivatives are contracted in pairs.
For a fixed $\triangle$, there are a finite number of independent operators, and the operator basis will be denoted by $\mathcal{B}^{\triangle}$.
This work is part of the efforts to find a basis of operators which are as simple and symmetric as possible.

\subsubsection*{Form factors}
The properties of a gluon operator $\mathcal{O}$ can be studied through its $n$-gluon form factors,
\begin{equation}
\mathcal{F}_n(\mathcal{O})\equiv
\int d^Dx e^{-iq\cdot x}\langle p_1\cdots p_n|\mathcal{O}(x)|0\rangle\ ,
\end{equation}
and the $L$-loop correction of $\mathcal{F}_n(\mathcal{O})$ will be denoted by $\mathcal{F}_n^{(L)}(\mathcal{O})$. 

The tree form factor $\mathcal{F}_n^{(0)}(\mathcal{O})=0$ if $n<N_F$, because each $F$ in $\mathcal{O}$ corresponds to at least one external leg through Wick's contraction.

\subsubsection*{Length and minimal form factor}
The $\mathbf{length}$ of an operator $\mathcal{O}$ is defined as the smallest integer $n$ for which the tree level form factor $\mathcal{F}_n^{(0)}(\mathcal{O})$ is non-vanishing, denoted by $\mathbb{L}(\mathcal{O})$.  While length is often informally interpreted as the number of $F$ terms in the operator, this naive counting can lead to ambiguity. A concrete example illustrating this ambiguity is provided by the following operators:
\begin{equation}
\begin{aligned}\label{equiva-op}
&\mathcal{O}_1=\Tr(D_{12}F_{34}F_{12}F_{34}),\ 
 \mathcal{O}_2=\Tr(D_{21}F_{34}F_{12}F_{34}),\ 
\\
&\mathcal{O}_1-\mathcal{O}_2=\Tr([D_{1},\ D_{2}]F_{34}F_{12}F_{34})
=ig_s\Tr([F_{12},F_{34}]F_{12}F_{34})\ .
\end{aligned}
\end{equation}
Here, both $\mathcal{O}_1$ and $\mathcal{O}_2$ are length-3 operators. However, their combination $\mathcal{O}_1-\mathcal{O}_2$ is a length-4 operator. 
Critically, the number of $F$ terms in $\mathcal{O}_1-\mathcal{O}_2$ varies depending on how the operator is algebraically rewritten, further highlighting the flaw in relying on raw $F$-counting to define length.

The length of a dimension-$\triangle$ operator satisfies $2\le \mathbb{L}(\mathcal{O})\le \frac{\triangle}{2}$, so the operator basis can be split into
\begin{equation}
\mathcal{B}^{\triangle}=\mathcal{B}^{\triangle}_{2}\cup\mathcal{B}^{\triangle}_{3}
\cup\cdots\cup \mathcal{B}^{\triangle}_{\frac{\triangle}{2}}\ ,
\end{equation}
in which $\mathcal{B}^{\triangle}_{i}$ is a set of length-$i$ operators.

While the total number of elements in the basis $\mathcal{B}^{\triangle}$ is fixed, the number of elements in each of its subsets $\mathcal{B}^{\triangle}_{i}$ depends on the choice of operators used to construct the basis.
To illustrate this dependence, we consider the dimension-8 operators $\mathcal{O}_1$, $\mathcal{O}_2$ and $\mathcal{O}_1-\mathcal{O}_2$ introduced in \eqref{equiva-op}.
When constructing the dimension-8 basis $\mathcal{B}^{8}$, we may freely select any two of these three operators for inclusion. Notably, the numbers of elements in the sub-bases $\mathcal{B}^{8}_3$ and $\mathcal{B}^{8}_4$ will differ depending on which pair of operators is chosen.

The $n$-gluon tree form factor with $n=\mathbb{L}(\mathcal{O})$ is the most simple non-zero form factor of $\mathcal{O}$, and it is called the \textbf{minimal form factor}.
\begin{equation}
\mathcal{F}^{\text{min}}(\mathcal{O})=\mathcal{F}_{\mathbb{L}(\mathcal{O})}(\mathcal{O})\ .
\end{equation}
If two operators of length-$\mathbb{L}$ differ by an operator of length greater $\mathbb{L}$, they share the same minimal form factor, and these two
operators will be called \textbf{equivalent operators}. For example, $\mathcal{O}_1$ and $\mathcal{O}_2$ in \eqref{equiva-op} are equivalent operators.

It is helpful to consider each equivalent class of operators as a whole. The minimal form factor serves as a one-to-one mapping from these equivalent classes to gauge invariant polynomials (with proper symmetries) of $p_i\cdot p_j$, $p_i\cdot \epsilon_j$, and color factors.
If we choose $\mathcal{B}^{\triangle}$ so that it contains one operator in each class of equivalent operators, then the number of elements in each $\mathcal{B}^{\triangle}_{\mathbb{L}}$ is fixed. 
With this choice, the operators with length higher than $\mathbb{L}$ cannot be written as linear combinations of operators in $\mathcal{B}^{\triangle}_{\mathbb{L}}$. So for each $\mathbb{L}$, $\mathcal{B}^{\triangle}_{\mathbb{L}}$ can be constructed independently. 
In fact, constructing $\mathcal{B}^{\triangle}_{\mathbb{L}}$ is equivalent to searching for a basis for $\mathbb{L}$-gluon minimal form factors with proper mass dimension. 

Minimal form factors are herein regarded as a representation of operators. This representation proves particularly useful for analyzing the linear relationships between operators, as the Bianchi identities, equations of motion (EOM), and gauge invariance are automatically satisfied within minimal form factors. A detailed discussion of this property is provided in Section \ref{subsection:constraint}.

\subsubsection*{Primary operators}

Some operators can be related to other operators with lower mass dimensions.
Suppose $\mathcal{O}_1=\partial ^2\mathcal{O}_2$, then $\mathcal{O}_1$ is called a \textbf{descendant} of $\mathcal{O}_2$. 
If an operator is not a descendant of any operator, then it is called a \textbf{primary} operator.

In each dimension $\triangle$, we only need to study the primary operators, because all physical informations of descendants can be directly obtained from their corresponding primary operators with lower dimensions. For example, the form factor of $\mathcal{O}_1$ and $\mathcal{O}_2$ are related by
\begin{equation}\label{primary-FF}
\mathcal{F}_n(\mathcal{O}_1)= Q^2\mathcal{F}_n(\mathcal{O}_2)\ ,
\end{equation}
in which $Q=p_1+\cdots +p_n$ is the total momenta of all external gluons.

A more general definition of descendant operators is that they are operators expressible as total derivatives or linear combinations of derivatives of other operators, including tensor operators.
For example, 
\begin{equation}
\mathcal{O}_{\mu\nu}=\partial_{\mu}\mathcal{O}_{1\nu}
+\partial_{\nu}\mathcal{O}_{2\mu}
\end{equation}
is also a descendant operator. 

\subsection{The kinematic operators}
\label{subsection:kinematic-operator}

A gluon operator can be split into a kinematic part and a color factor.  For example, for the operator  $\mathcal{O}=\Tr(F_{12}D_{1}F_{34}D_{2}F_{34})$,
the kinematic part is $(F_{12})^{a_1}(D_{1}F_{34})^{a_2}(D_{2}F_{34})^{a_3}$, and the color factor is $\Tr(T^{a_1}T^{a_2}T^{a_3})$.

The kinematic part of an operator will be called a \textbf{kinematic operator}. A kinematic operator can be split into several blocks, and each block carries a color index $a_i$. For example, $ (F_{12})^{a_1}(D_{1}F_{34})^{a_2}(D_{2}F_{34})^{a_3}$ has three blocks
\begin{equation}\label{ki-example}
\mathtt{k}_1^{a_1}=(F_{12})^{a_1},\ 
\mathtt{k}_2^{a_2}=(D_{1}F_{34})^{a_2},\ 
\mathtt{k}_3^{a_3}=(D_{2}F_{34})^{a_3}.\ 
\end{equation}
$\mathtt{k}_i$ will be called  \textbf{DF-blocks}. 

We will also use the notation
\begin{equation}
\lfloor\mathtt{k}_1,\mathtt{k}_2,\cdots,\mathtt{k}_n\rfloor
\equiv \mathtt{k}_1^{a_1}\mathtt{k}_2^{a_2}\cdots \mathtt{k}_n^{a_n}\ .
\end{equation}
Then a length-$n$ kinematic operator $K$ has the form:
\begin{equation}
K=\lfloor\mathtt{k}_1,\mathtt{k}_2,\cdots,\mathtt{k}_n\rfloor\ .
\end{equation}
For compectness, sometimes we use the simplified notation,
\begin{equation}\label{k-o-define-2}
K\rightarrow  \mathtt{k}_1\mathtt{k}_2\cdots\mathtt{k}_n\ ,
\end{equation}
but we must keep in mind in this notation different DF-blocks do not commute with each other.
For example, 
\begin{equation}
\begin{aligned}
&F_{12}D_{1}F_{34}D_{2}F_{34}
\equiv\lfloor F_{12},D_{1}F_{34},D_{2}F_{34}\rfloor
=(F_{12})^{a_1}(D_{1}F_{34})^{a_2}(D_{2}F_{34})^{a_3}\ ,\\
&D_{1}F_{34}F_{12}D_{2}F_{34}
\equiv\lfloor D_{1}F_{34},F_{12},D_{2}F_{34}\rfloor
=(D_{1}F_{34})^{a_1}(F_{12})^{a_2}(D_{2}F_{34})^{a_3}\ .\\
\end{aligned}
\end{equation}

We will define the \textbf{kinematic form factor} for a kinematic operator $K$ as
\begin{equation}\label{kinematic-ff-1}
\mathtt{F}(K)\equiv\prod_{i=1}^n \mathtt{k}_i(p_i,\epsilon_i)
\equiv\prod_{i=1}^n \mathtt{k}_i
\Bigr|_{D_{\mu}\rightarrow p_{i\mu},\ F_{\mu\nu}\rightarrow p_{i\mu}\epsilon_{i\nu}-p_{i\nu}\epsilon_{i\mu} }\ .
\end{equation}
From \eqref{kinematic-ff-1} we can see that kinematic form factors are gauge invariant:
\begin{equation}
\mathtt{F}(K)\Bigr|_{\epsilon_i\rightarrow p_i}=0\ .
\end{equation}
In fact, any quantity $Y$ satisfying the following conditions can be regarded as the kinematic form factor of a certain kinematic operator:
\begin{enumerate}
\item $Y$ is a polynomial of Mandelstam variables $s_{ij}\equiv 2p_i\cdot p_j$, $p_i\cdot \epsilon_j$ and $\epsilon_i\cdot \epsilon_j$.

\item $Y$ is multilinear in each $\epsilon_i$.

\item $Y$ is gauge invarint. 
\end{enumerate}
The corresponding kinematic operator can be denoted by $\mathtt{F}^{-1}(Y)$.

DF-blocks and the kinematic form factor can be regarded as two different representations of a kinematic operator. 
The DF-block representation is more compact, but in this form it can be difficult to study relations like EOM and Bianchi identities. However, as will be discussed in Section \ref{subsection:constraint}, these relations are automatically satisfied in the kinematic form factor representation.
In Section \ref{section:diagram}, we will introduce a graphical represenation, which will be the main focus of this paper.

Our basis of dimension-$\triangle$ kinematic operators will be denoted by $\mathcal{K}^{\triangle}$. Similar as the color-dressed operators, the basis can also be split into several subsets according to the operator length:
\begin{equation}
\mathcal{K}^{\triangle}=\mathcal{K}^{\triangle}_2\cup\cdots 
\cup \mathcal{K}^{\triangle}_{\frac{\triangle}{2}}\ .
\end{equation}

If a kinematic operator $K$ satisfies 
\begin{equation}\label{define-weight}
\mathtt{F}(K)|_{p_i\rightarrow \lambda p_i}=\lambda^{w_i} \mathtt{F}(K),\ i=1,\cdots, n,
\end{equation}
then $K$ is called a \textbf{homogeneous operator} with weight $\textbf{w}=(w_1,\cdots, w_n)$. 
For example, the kinematic operator given by \eqref{ki-example} is a homogeneous operator with weight $(1,2,2)$.
Weight can also be defined for Mandelstam variables. For example, $\textbf{w}(s_{12})=(1,1,0,\cdots,0)$.

Each $\mathcal{K}^{\triangle}_{\mathbb{L}}$ can be split into several subsets with fixed weight. As an example,
\begin{equation}
\mathcal{K}^{8}_{3}=\mathcal{K}^{113}_{3}\cup \mathcal{K}^{131}_{3}\cup \mathcal{K}^{311}_{3}\cup \mathcal{K}^{122}_{3}\cup \mathcal{K}^{212}_{3}\cup \mathcal{K}^{221}_{3}\ ,
\end{equation}
in which $\mathcal{K}^{113}_{3}$ contains the operators with weight $(1,1,3)$.
If a weight $\textbf{w}_2$ can be obtained from another  weight $\textbf{w}_1$ by permutations, then $\mathcal{K}^{\textbf{w}_2}_{\mathbb{L}}$ can also be obtained from $\mathcal{K}^{\textbf{w}_1}_{\mathbb{L}}$ by permutations.

\subsection{The constraints on DF-blocks}
\label{subsection:constraint}
A combination of kinematic operators in DF-block form vanishes if it is proportional to the equation of motion (EOM), Bianchi identities. Besides, a kinematic operator can be reduced to higher length operators if it contains $D_{\mu}$ commutators:
\begin{equation}\label{constraints-3}
D_1F_{12}=0\ ,
D_1F_{23}+D_2F_{31}+D_3F_{12}=0,\ 
[D_1,D_2]=ig_s F_{12}\ .
\end{equation}
This means kinematic operators may not be independent, even if they have different DF-block forms.

In the Language of kinematic form factors, \eqref{constraints-3} are automatically satisfied:
\begin{equation}
\begin{aligned}
&D_1F_{12}=0 &\Rightarrow&
 p^{\mu}(p_{\mu}\epsilon_{\nu}-p_{\nu}\epsilon_{\mu})=0,\ \\
&D_{[1}F_{23]}=0 &\Rightarrow &p_{[\mu}p_{\nu}\epsilon_{\rho]}=0,\ \\
&[D_{1},D_{2}]=ig_s F_{12}&\Rightarrow &[p_{\mu},p_{\nu}]=0\ .\\
\end{aligned}\label{constraints-3-1}
\end{equation}
So in order to construct the basis of kinematic operators, $\mathcal{K}^{\textbf{w}}_{\mathbb{L}}$, we may first enumerate all different kinematic operators with the weight $\textbf{w}$, then select independent ones by examing their kinematic form factors.
However, there are usually a large number of different kinematic operators, because there are numerous ways to contract the Lorentz indices carried by $F$ and $D$.
It would be desirable to remove as many redundant operators as possible before we start.


Using \eqref{constraints-3}, we can find some patterns of redundant operators, and derive the following rules which can be used to constrain the DF-blocks:

\subsubsection*{\textbf{DF rule 1}: $D$-indices in the same DF-block commute with each other. }
\label{DF rule 1}
Lorentz indices can be carried by $F$ or $D$. If an index is carried by $F$($D$), we will call it a $F$-index($D$-index). An index appears twice in an operator, and it can be a $F$-index in one block, and a $D$-index in the other block.
A pair of contracted $D$-index will be called a $\mathbf{D}$\textbf{-pair}.

The commutator of two $D$'s produce an $F$, $[D_{1},\ D_{2}]X=ig_s[F_{12},X]$, so the permutation of of $D$'s in a DF-block only change the operator by some higher length terms:
\begin{equation}
D_1D_2\cdots D_nF_{ab}=D_{\sigma_1}\cdots D_{\sigma_n}F_{ab}+\mathcal{O}(F^2)\ ,
\end{equation}
in which $(\sigma_1,\cdots,\sigma_n)$ is a permutation of $(1,\cdots,n)$.
Therefore DF-blocks which are related by permutations of $D$-indices can be regarded as equivalent.
We will also use the abbreviation $D_{12\cdots n}\equiv D_1D_2\cdots D_{n}$.

\subsubsection*{\textbf{DF rule 2}: 
All Lorentz indices in the same DF-block must be different. }
\label{DF rule 2}

In a DF-block, two $F$-indices must be different because $F_{\mu\nu}$ is anti-symmetric. Each $F$-index must be different from all $D$-indices, otherwise the operator vanishes using EOM. Also, a $D$-pair in the same block increases the length of the operator,
\begin{equation}
\begin{aligned}
D^2F_{12}=&D_3D_3F_{12}=-D_3D_1F_{23}+D_3D_2F_{13}\\
=&-[D_3,D_1]F_{23}+[D_3,D_2]F_{13}
\sim \mathcal{O}(F^2)\ ,\\
\end{aligned}
\end{equation}
in which we used EOM and Bianchi identities.

From now on, we will assume all Lorentz indices in the same DF-block are different, unless otherwise specified.

\subsubsection*{\textbf{DF rule 3}: When starting with a DF-block with $n$ Lorentz indices, $n-1$ independent DF blocks can be generated by permuting these Lorentz indices. }
\label{DF rule 3}

If we permute the Lorentz indices in $D_1F_{23}$, three DF-blocks can be generated: $D_1F_{23}$, $D_2F_{31}$ and $D_3F_{12}$. But only two of them are independent because of  Bianchi identity.

DF-blocks with $n$ Lorentz indices have $n-2$ $D$-indices and two $F$-indices. They have the form:
\begin{equation}\label{fij-define}
f_{ij}\equiv D_{1,\cdots, i-1,i+1,\cdots, j-1,j+1,\cdots, n}F_{ij}\ .
\end{equation}
$f_{ij}$ can be rewritten as $f_{i,j-1}+f_{j-1,j}$ using Bianchi identies:
\begin{equation}
\begin{aligned}
&f_{ij}
\sim D_{1,\cdots, j-2,j+1,\cdots, n}F_{j-1,j}
+D_{1,\cdots, i-1,i+1,\cdots ,j-2,j\cdots n}F_{i,j-1}
=f_{i,j-1}+f_{j-1,j}\ ,\\
\end{aligned}\label{fbianchi}
\end{equation}
in which we have dropped higher length terms. Using \eqref{fbianchi} repeatedly we derive
\begin{equation}
f_{ij}\sim \sum_{k=i}^{j-1} f_{k,k+1}\ .
\end{equation}
So there are $n-1$ independent DF-blocks, $\{f_{i,i+1}|i=1,\cdots, n-1\}$. 

\subsection{Primitive operators}
\label{subsection:primitive}



The definition of primary and descendant operators can be easily extended to kinematic operators satisfying $K_1=\partial^2K_2$. The kinematic form factors of $K_1$ and $K_2$ satisfy the same relation as \eqref{primary-FF},
\begin{equation}\label{primary-FF-1}
\mathtt{F}(K_1)=(p_1+\cdots +p_{\mathbb{L}})^2\mathtt{F}(K_2)\ ,
\end{equation}
in which $\mathbb{L}$ is the length of $K_1$ and $K_2$.

Descendant kinematic operators (e.g. $K_1$) can be derived directly from lower-dimensional kinematic operators (e.g. $K_2$) via \eqref{primary-FF-1}, and one only need to construct the primary kinematic operators.  Consequently, the task of building an operator basis can be simplified using relations that link higher-dimensional operators to lower-dimensional ones. We focus on the following such relation:
\begin{equation}
\mathtt{F}(K_1)=p_i\cdot p_j\mathtt{F}(K_2).
\end{equation}
Here, $K_1$ is obtained by adding a $D$-pair to $K_2$, where this $D$-pair connects the $i$-th and $j$-th DF-blocks. To illustrate, consider $K_1=F_{12}D_4F_{23}D_4F_{31}$ and $K_2=F_{12}F_{23}F_{31}$. Their form factors satisfy the relation $\mathtt{F}(K_1)=p_2\cdot p_3\mathtt{F}(K_2)$.

Kinematic operators can be divided into two classes based on whether they contain $D$-pairs. Operators with $D$-pairs will be called \textbf{non-primitive operators}. 
Independent operators without $D$-pairs will be called \textbf{primitive operators}. Non-primitive operators can be generated from primitive operators systematically by adding $D$-pairs.

For example, the only length-2 primitive operator is $F_{12}F_{12}$, which has dimension-4. The dimension-6 non-primitive operators can be obtained by adding a $D$-pair to it:
\begin{equation}
\mathcal{K}^6_2=\{D_3F_{12}D_3F_{12}\}\ .
\end{equation}
Similarly, the non-primitive operators with dimension $\triangle=2n$ is \cite{Jin:2020pwh}
\begin{equation}
\mathcal{K}^{2n}_2=\{D_{3\cdots n}F_{12}D_{3\cdots n}F_{12}\}\ .
\end{equation}

More precisely, let $L^{\triangle}_{\mathbb{L},np}$ be the linear space spanned by dimension-$\triangle$ length-$\mathbb{L}$ non-primitive operators, then the space spanned by dimension-$\triangle$ length-$\mathbb{L}$ operators, $\text{span}(\mathcal{K}_{\mathbb{L}}^{\triangle})$, can be decomposed into the direct sum of $L^{\triangle}_{\mathbb{L},np}$ and its quotient space, denoted by $L^{\triangle}_{\mathbb{L},p}$:
\begin{equation}\label{prim-subspace}
\text{span}(\mathcal{K}_{\mathbb{L}}^{\triangle})
=L^{\triangle}_{\mathbb{L},np}\oplus L^{\triangle}_{\mathbb{L},p}\ .
\end{equation}
The primitive operator basis, $\mathcal{P}^{\triangle}_{\mathbb{L}}$, is the basis of $L^{\triangle}_{\mathbb{L},p}$,
\begin{equation}
L^{\triangle}_{\mathbb{L},p}=\text{span}(\mathcal{P}^{\triangle}_{\mathbb{L}})\ .
\end{equation}

It is possible that although two operators $P_1$ and $P_2$ contain no $D$-pair, but $P_1+P_2\in L^{\triangle}_{np}$. For example,
\begin{equation}
\begin{aligned}
P_1=&F_{12}F_{13}D_2F_{45}D_4F_{36}F_{56}\ ,\\
P_2=&F_{12}F_{13}D_5F_{24}D_4F_{36}F_{56}\ ,\\
P_1+P_2=&F_{12}F_{13}D_4F_{25}D_4F_{36}F_{56}\ ,\\
\end{aligned}\label{prim-to-non}
\end{equation}
in which we used Bianchi identity.
In \eqref{prim-subspace}, since $P_1+P_2\in L^{\triangle}_{\mathbb{L},np}$,  $P_1$ and $P_2$ cannot be in the primitive operators basis at the same time. More formally, the primitive operator $P_1$ can be identified as the equivalence class $\{P_1+K|K\in L^{\triangle}_{\mathbb{L},np}\}$.

There can be different ways to choose primitive operator basis. 
However, the number of independent primitive operators is fixed:
\begin{equation}
\text{dim}\Bigl[L^{\triangle}_{\mathbb{L},p}\Bigr]
=\text{dim}\Bigl[\text{span}(\mathcal{K}_{\mathbb{L}}^{\triangle})\Bigr]
-\text{dim}\Bigl[L^{\triangle}_{\mathbb{L},np}\Bigr]\ .
\end{equation}

As will be shown later, for a given length, the dimension of primitive operators has a upper bound, and there are only a finite number of independent primitive operators.
At length-3, there are only 4 independent primitive operators \cite{Jin:2020pwh},
\begin{equation}
\begin{aligned}
&\mathcal{P}^6_3=\{F_{12}F_{23}F_{31}\}\ ,\\
&\mathcal{P}^8_3=\{F_{12}D_1F_{34}D_2F_{34}\ ,
D_2F_{34}F_{12}D_1F_{34}\ ,
D_1F_{34}D_2F_{34}F_{12}\},\ \\
\end{aligned}\label{length-3-prim}
\end{equation}
in which the 3 operators in $\mathcal{P}^8_3$ are related by permutations of DF-blocks.
From them length-3 kinematic operators with arbitrary dimension can be easily constructed. For example
\begin{equation}
\begin{aligned}
\mathcal{K}^6_3=&\{F_{12}F_{23}F_{31}\}\ ,\\
\mathcal{K}^8_3=&\{F_{12}D_4F_{23}D_4F_{31},\ F_{12}D_1F_{34}D_2F_{34}\}
\cup\text{permutations of DF-blocks}\ ,\\
\mathcal{K}^{10}_3=&\{F_{12}D_{45}F_{23}D_{45}F_{31},\ D_4F_{12}D_5F_{23}D_{45}F_{31},\ 
F_{12}D_{15}F_{34}D_{25}F_{34},\ \\
&\ \ D_5F_{12}D_1F_{34}D_{25}F_{34}\}\cup\text{permutations of DF-blocks}\ .\\
\end{aligned}
\end{equation}

The primitive operators with higher length can be constructed systematically using the operator diagram technique which will be introduced in the next section.

\section{Graphical representation and primitive operators}
\label{section:diagram}

DF-blocks and kinematic form factors can be regarded as two different representations of kinematic operators.
In this section we propose a new graphical representation, which allows us to carry out a systematical construction of primitive operators.
First we introduce operator diagrams, in which DF-blocks, Lorentz indices and covariant derivatives are represented by vertices, edges and arrows, respectively.
Then we discuss primitive configurations, which are primitive operators in the language of operator diagrams. 

\subsection{Operators diagrams}
\label{subsection:op-diagram}

The kinematic operators can be converted to diagrams using the following rules:
\begin{enumerate}
\item Each DF-block corresponds to a vertex of the diagram. 

\item Each pair of Lorentz indices are connected together to form an edge of the diagram.

\item If $\mu$ is a $D$-index in the vertex $\mathtt{k}_i$, then we add an arrow to the edge $\mu$ pointing to $\mathtt{k}_i$.

\end{enumerate}


\begin{figure}
\centering
\includegraphics[scale=0.9]{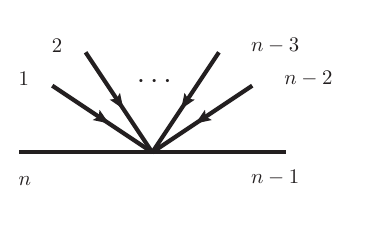}
\caption{A vertex representating the DF-block $D_{12\cdots(n-2)}F_{n-1,n}$.}
\label{fig:vertex-define}
\end{figure}
The DF-block $D_{12\cdots (n-2)}F_{n-1,n}$ is represented by the vertex shown in Figure \ref{fig:vertex-define}. The edges labelled with 1 to $n-2$ are $D$-indices, and they have arrows coming into the vertex. Note that the sign of the operator changes when reordering the $F$-indices, for instance $D_{12\cdots (n-2)}F_{n,n-1}=-D_{12\cdots (n-2)}F_{n-1,n}$. In the operator diagram formalism, however, we omit explicit consideration of this sign ambiguity. The correct sign will be explicitly derived from the corresponding DF-form when required.

\begin{figure}
\centering
\includegraphics[scale=0.8]{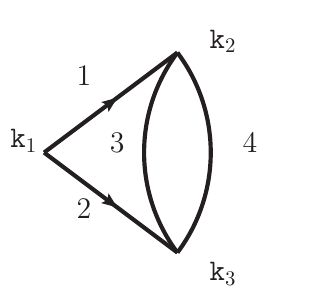}
\caption{The operator diagram of $F_{12}D_1F_{34}D_2F_{34}$.}
\label{fig:op-dia-ex-1}
\end{figure}

For example, the operator diagram of $F_{12}D_1F_{34}D_2F_{34}$ in \eqref{length-3-prim} is shown in Figure \ref{fig:op-dia-ex-1}. The operator has 3 DF-blocks and 4 pairs and contracted Lorentz indices. Correspondingly, the diagram contains 3 vertices $\mathtt{k}_1$, $\mathtt{k}_2$, $\mathtt{k}_3$, and 4 edges labelled with $1,2,3,4$.

\begin{figure}
\centering
\includegraphics[scale=0.5]{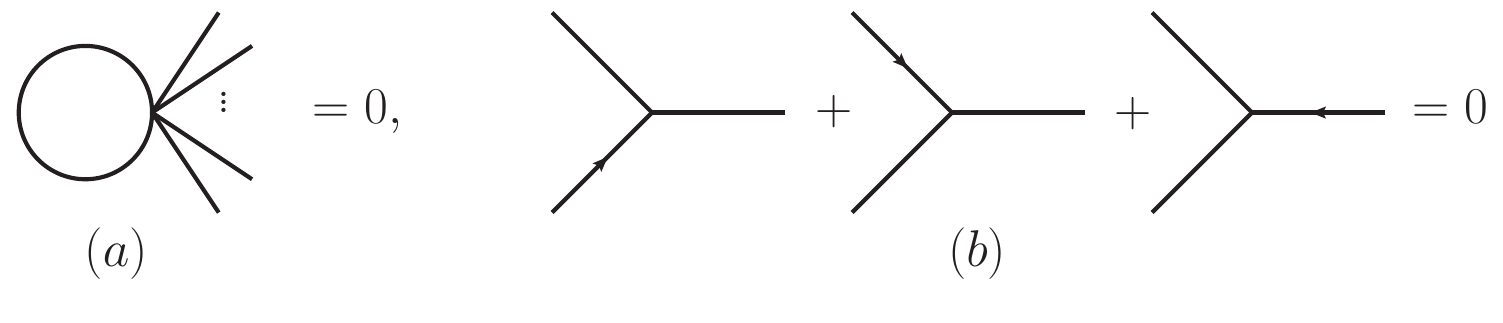}
\caption{ ($a$): \textbf{DF rule 2} and  ($b$): Bianchi identity in operator diagram language.}
\label{fig:tadpole1}
\end{figure}

The number of vertices in an operator diagram equals to the length of the operators,  $V=\mathbb{L}$. The number of edges $E=\mathbb{L}+\frac{N_D}{2}=\frac{\triangle}{2}$. The degree of a vertex $v$, denoted by $\text{deg}(v)$, refers to the number of edges which are adjacent to $v$. 

The properties of kinematic operators in Section \ref{subsection:constraint}  can be translated from the DF-block language to the operator diagram language. For example, \textbf{DF rule 1} on page \pageref{DF rule 1} means that the directed edges adjacent to a vertex are unordered. \textbf{DF rule 2} is translated to "two endpoints of an edge cannot be connected to the same vertex", as shown in Figure \ref{fig:tadpole1}(a). The Bianchi identity is translated to "three arrow configurations of a 3-vertex add up to zero" as shown in Figure \ref{fig:tadpole1}(b).

\begin{figure}
\centering
\includegraphics[scale=0.7]{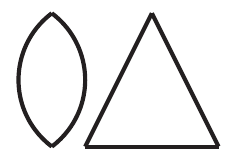}
\caption{A disconnected  operator diagram.}
\label{fig:disconnected-diagram}
\end{figure}

Operator diagrams can be connected or disconnected.  An example of disconnected diagram is given in Figure \ref{fig:disconnected-diagram}, and it corresponds to the operator $F_{12}F_{12}F_{34}F_{45}F_{53}$. The operator can be divided into two components, $F_{12}F_{12}$ and $F_{34}F_{45}F_{53}$, and there are no  Lorentz contractions between these two components. We will henceforth assume the operator diagrams are connected (unless stated otherwise), as the components can be treated separately.

\subsection{Arrow configurations}
\label{subsection:arrow-config}

If we remove all arrows from an operator diagram, the resulting diagram will be called an \textbf{undirected diagram}. 
In the terminology of Feynman diagrams, the undirected diagrams are vacuum diagrams without external legs.\footnote{Diagrams with external legs correspond to tensor operators with uncontracted Lorentz indices.} 
There are usually multiple ways to add arrows to an undirected diagram (or any subdiagram of an  undirected diagram), and we will call them (arrow) \textbf{configurations}.

\label{vertex-config-ext}
Let us start with the arrow configuration of a vertex. The arrows on the edges adjacent to a vertex can be incoming or outgoing, but only the incoming arrows represents $D_{\mu}$ acting on the vertex. So the arrow configurations of a vertex is defined by the incoming arrows on the edges adjacent to the vertex. 
The arrow configurations of an operator diagram can be obtained by combining the arrow configurations of all its vertices.

\begin{figure}
\centering
\includegraphics[scale=0.7]{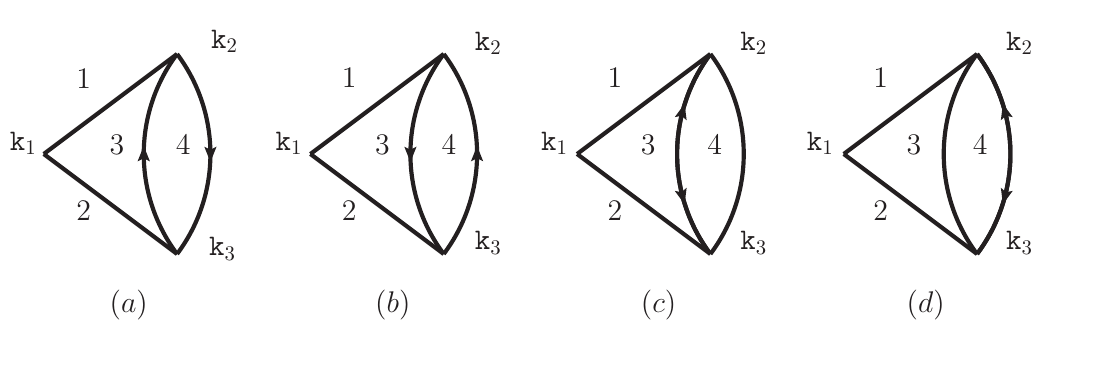}
\caption{Different arrow configurations of a diagram.}
\label{fig:arrow-config}
\end{figure}

Different arrow configurations correspond to different kinematic operators. For example, the configurations in  Figure \ref{fig:arrow-config} correspond to the following operators:\footnote{Figure \ref{fig:op-dia-ex-1} can be written as a linear combination of 4 configurations in Figure \ref{fig:arrow-config} using Bianchi identity.}
\begin{equation}
\begin{aligned}
&(a): F_{12}D_3F_{14}D_4F_{23}\ ,\\
&(b): F_{12}D_4F_{13}D_3F_{24}\ ,\\
&(c): F_{12}D_3F_{14}D_3F_{24}\ ,\\
&(d): F_{12}D_4F_{13}D_4F_{23}\ .\\
\end{aligned}
\end{equation}



\subsubsection*{Symmetries of operator diagrams}

Relabelling the vertices $\mathtt{k}_i$ corresponds to a permutation of the DF-blocks; accordingly, diagrams that differing solely by such vertex relabelling are regarded as distinct operator diagrams, unless the diagrams are invariant under this relabelling. For examle, relabelling $\mathtt{k}_1\leftrightarrow \mathtt{k}_2$ in Figure \ref{fig:arrow-config}(a) yields a different operator $D_3F_{14}F_{12}D_4F_{23}$. By contrast, relabelling $\mathtt{k}_2\leftrightarrow \mathtt{k}_3$ leaves the operator unchanged.

Conversely, two operator diagrams are regarded as the same if they can be transformed into each other by relabelling the edges, as the edges correspond to dummy Lorentz indices. Such equivalence is only feasible when the diagram contains bubble-like structures. For instance, in Figure \ref{fig:arrow-config}, diagrams (a) and (b), as well as (c) and (d), are related by the edge relabelling bubble edges $3\leftrightarrow 4$, and thus represent identical operator diagrams.

The symmetry group of a undirected diagram can be written as
\begin{equation}
G=G_T\times G_B\ ,
\end{equation}
in which $G_T$ is generated by permutation symmetry of vertices, while $G_B$ is generated by the permutation symmetry of bubble edges. For example, for the graph in Figure \ref{fig:disconnected-diagram}, $G_T=S_3\times S_2$ and $G_B=S_2$.

The analysis of symmetries can be applied to operator diagrams with and without arrow configurations.

\subsubsection*{Independent arrow configurations}

Arrow configurations of a diagram are generally not independent due to relations imposed by the Bianchi identity. Suppose $T$ is an undirected diagram with vertices $v_1,\cdots ,v_n$.  The independent configurations of $T$ can be obtained by combining the independent configurations of each vertex. For each vertex $v_i$, there are $C_{\text{deg}(v_i)}^2$ possible configurations, but only $\text{deg}(v_i)-1$ are independent, as specified by \textbf{DF rule 3} on page \pageref{DF rule 3}.

\begin{figure}
\centering
\includegraphics[scale=0.6]{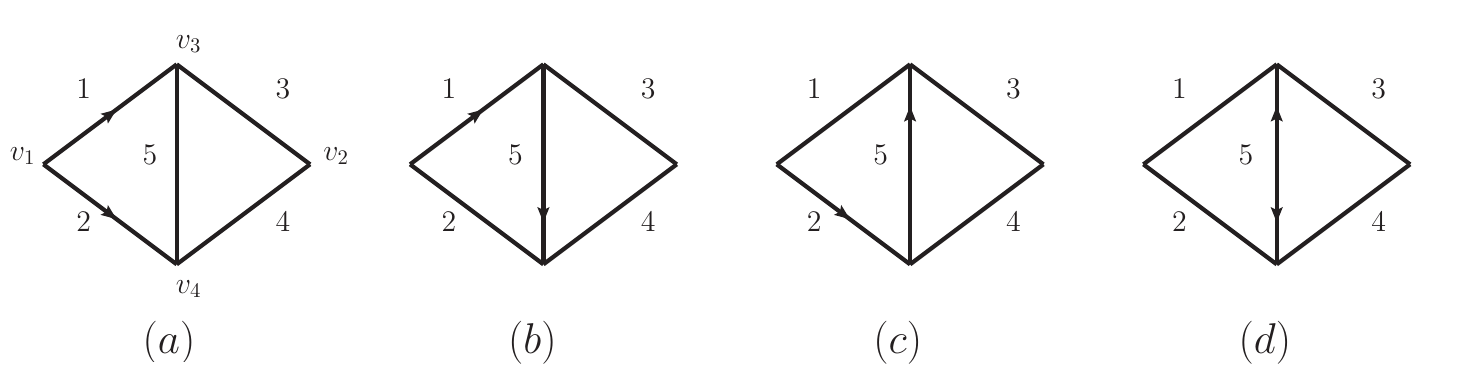}
\caption{Independent configuations of a vertices are combined into independent configurations of the diagram.}
\label{fig:vertex-config}
\end{figure}

As an example, in Figure \ref{fig:vertex-config}, the vertices $v_3$ and $v_4$ each have degree 3, so both of them have a single arrow coming in. Using the Bianchi identity, $v_3$ has 2 independent arrow configurations, and we choose to add arrow to edge-1 or edge-5. Similarly, we add arrow to edge-2 or edge-5 for the vertex $v_4$. $v_1$ and $v_2$ have no arrows coming in because they have degree 2. Therefore the undirected diagram has 4 independent configurations corresponding to the following operators:
\begin{equation}
\begin{aligned}
&(a): F_{12}F_{34}D_1F_{35}D_2F_{45}\ ,\\
&(b): F_{12}F_{34}D_1F_{35}D_5F_{24}\ ,\\
&(c): F_{12}F_{34}D_5F_{13}D_2F_{45}\ ,\\
&(d): F_{12}F_{34}D_5F_{13}D_5F_{24}\ .\\
\end{aligned}\label{config-4-DF}
\end{equation}

However, extra attention must be paid to digrams with bubble structures, for which not all configurations are independent. A bubble is formed by two vertices connected by two or more edges. For example, the diagram in Figure \ref{fig:op-dia-ex-1} has a bubble because $\mathtt{k}_2$ and $\mathtt{k}_3$ are connected by edge-3 and edge-4. Two configurations should be regarded as the same if they are related by a permutation of bubble edges.  Therefore Figure \ref{fig:arrow-config}(a) and \ref{fig:arrow-config}(b) should be regarded as the same configuration, and the same is true for Figure \ref{fig:arrow-config}(c) and \ref{fig:arrow-config}(d). This means Figure \ref{fig:arrow-config} only has two independent configurations.

The total number of independent configurations of a diagram with $G=G_T\times G_B$ is given by
\begin{equation}\label{n-config}
N_{\text{config}}(T)=\frac{1}{|G_B|}\prod_{i=1}^n[\text{deg}(v_i)-1].
\end{equation}

\subsection{Primitive configurations}
\label{subsection:prim-config}
If an arrow configuration corresponds to a primitive or non-primitive operator, we refer to it as a primitive or non-primitive configuration, respectively.
In the operator diagram language, a $D$-pair is an edge with two arrows. For example, Figure \ref{fig:vertex-config}(d) is a non-primitive configuration, since it correponds to the operators \eqref{config-4-DF}(d), which contains a $D$-pair corresponding to edge-5. 

\begin{figure}
\centering
\includegraphics[scale=0.6]{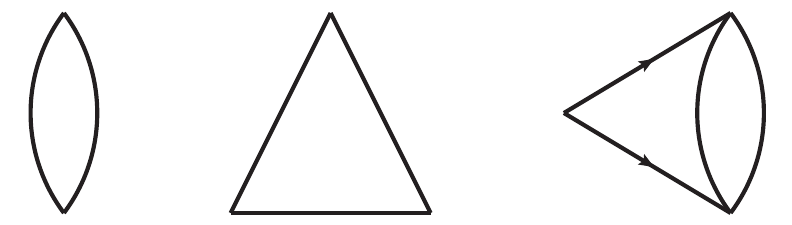}
\caption{Length-2 and length-3 primitive operator diagrams.}
\label{fig:length-2-3}
\end{figure}

It is very convenient to construct  primitive operators in the language of operator diagrams. 
Since Bianchi identities only relate different configurations of the same undirected diagram, it is clear that we can search for primitive configurations for each undirected diagram separately. All primitive configurations of length 2 and length 3 can be derived from those shown in Figure \ref{fig:length-2-3} through vertex permutations.

\begin{figure}
\centering
\includegraphics[scale=0.7]{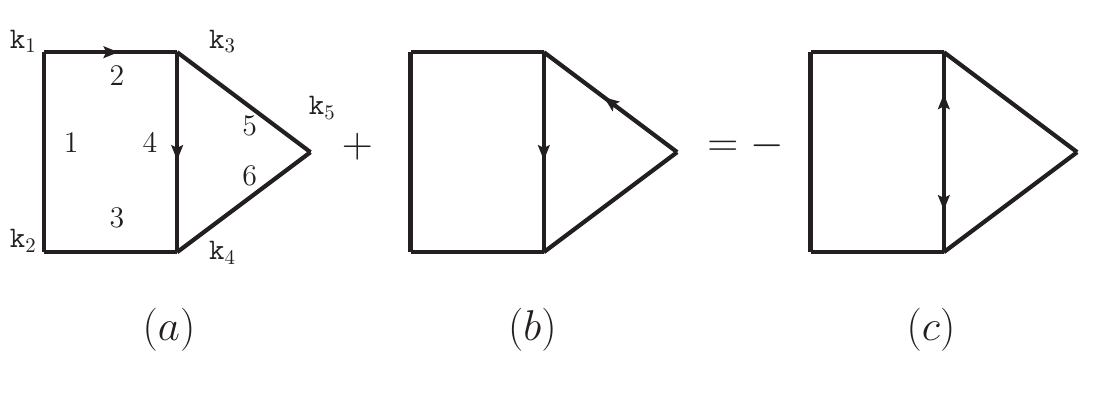}
\caption{Two primitive configurations sum to a non-primitive configuration.}
\label{fig:np-combination}
\end{figure}

However, for more complex diagrams, independent primitive configurations cannot be obtained merely by removing all configurations that contain $D$-pairs. As shown in \eqref{prim-to-non}, a comination of configurations without $D$-pairs may become non-primitive. The diagrammatic representation of \eqref{prim-to-non} is given in Figure \ref{fig:np-combination}, in which Figure \ref{fig:np-combination}(a) and (b) are two configurations without $D$-pairs, but their sum can be converted into a non-primitive configuration Figure \ref{fig:np-combination}(c) using Bianchi identity.

It is therefore necessary to develop a more systematic method for obtaining independent primitive configurations. Before analyzing the primitive configurations of complete diagrams, however, we first examine those of simpler substructures or sub-diagrams.
A sub-diagram of a undirected diagram $T$ comprises of a subset of $T$'s vertices along with all their adjacent edges. 

We start with the sub-diagram $T_{v_1v_2}$ which consists of $v_1$, $v_2$ and all edges adjacent to them. The configurations of this sub-diagram\footnote{As discussed on page \pageref{vertex-config-ext}, when specifying the configuration of a vertex, only the directions of the incoming arrows are considered. Similarly, for a sub-diagram, only the incoming arrows on the external legs are specified when defining its configuration.} can be obtained by combining the configurations of $v_1$ and $v_2$. Some of these configurations can be non-primitive. Following the discussion in Section \ref{subsection:primitive}, we can decompose the linear space spanned by all configurations to a non-primitve subspace $L_{np}$ and its quotient space $L_p$, and the basis of $L_p$ gives the independent primitive configurations of $T_{v_1v_2}$.
We can continue to add more vertices to the sub-diagram and remove the non-primitive configurations, until all vertices are included in the sub-diagram, and the primitive configurations of the complete diagram are obtained.

Bianchi identities are only allowed for $\text{deg}\ge 3$ vertices, so deg-2 vertices usually play no role in this process, except for a special case which will be discussed later. We can divide the vertices into two classes, deg-2 vertices and $\text{deg}\ge 3$ vertices. Then edges can be divided into 3 types:
\begin{enumerate}[I]
\item Type-22: Both endpoints are deg-2 vertices. No arrow is allowed on the edge. For example, edge 1 in Figure \ref{fig:np-combination}(a).

\item Type-2v: One endpoint is a deg-2 vertex, and the other is a $\text{deg}\ge 3$ vertex. At most one arrow is allowed on the edge. For example, edges 2,3,5,6 in Figure \ref{fig:np-combination}(a).

\item Type-vv: Both endpoints are $\text{deg}\ge 3$ vertices. At most two arrows are allowed on the edge. For example, edge 4 in Figure \ref{fig:np-combination}(a).

\end{enumerate}
$D$-pairs are only allowed on type-vv edges.

As an example, in Figure \ref{fig:np-combination}, edge-1 is  a type-22 edge, edge-2, 3, 5, 6 are type-2v edges, and edge-4 is a type-vv edge.

Clearly, type-22 edges do not play a role in the construction of primitive configurations. By removing the degree-2 vertices and type-22 edges from $T$, we obtain a subdiagram which will be called the \textbf{skeleton diagram}, and denoted by $T_0$. Since all arrows in the diagram are attached to the skeleton diagram, each primitive configuration of the skeleton diagram corresponds to a primitive configuration of the original diagram.

The internal legs of the skeleton diagram are type-vv edges of the original operator diagram, and the external legs of skeleton diagrams are type-2v edges. The number of external legs will be denoted by  $\mathcal{E}(T_0)$. 

\begin{figure}
\centering
\includegraphics[scale=0.7]{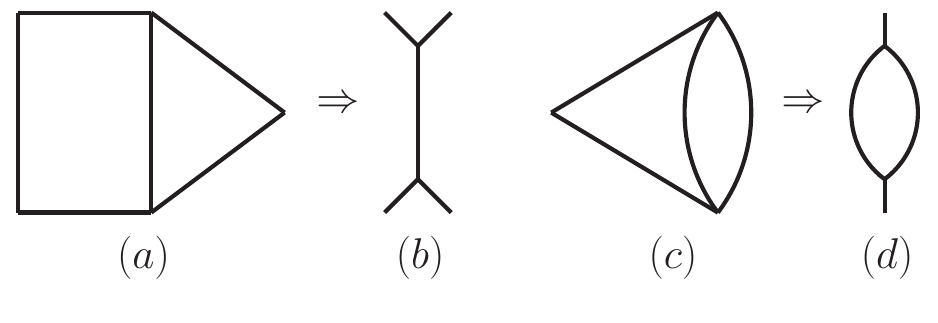}
\caption{Skeleton diagrams.}
\label{fig:non-prim-sub}
\end{figure}

The skeleton diagram of Figure \ref{fig:np-combination} is given in Figure \ref{fig:non-prim-sub}(b). The skeleton diagram of Figure \ref{fig:op-dia-ex-1} is given in Figure \ref{fig:non-prim-sub}(d).

Different operator diagrams may have the same skeleton diagram. For example, the skeleton diagram of Figure \ref{fig:vertex-config} is the same as Figure \ref{fig:non-prim-sub}(b).
So by constructing the primitive configurations of a skeleton diagram $T_0$, we actually obtain the primitive configurations of all undirected diagrams whose skeleton diagrams are the same as $T_0$.

\begin{figure}
\centering
\includegraphics[scale=0.7]{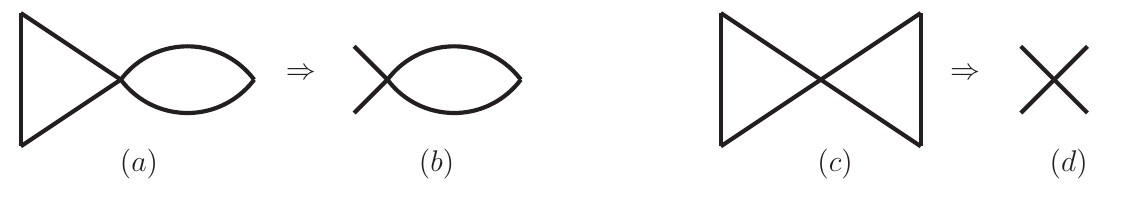}
\caption{The skeleton snail is considered as part of the skeleton diagram.}
\label{fig:np-loop}
\end{figure}

A special structure which requires extra attention is the 1-loop bubble with a deg-2 vertex (see e.g. Figure \ref{fig:np-loop}(a)). The structure will be called a \textbf{skeleton snail}.
We will add this deg-2 vertex to the skeleton diagram, because otherwise we cannot discriminate the skeleton diagrams of Figure \ref{fig:np-loop}(a) and \ref{fig:np-loop}(c). These two diagrams have different number of primitive configurations, because the former has a non-trivial $G_B=S_2$.

We will use $\mathcal{R}(T)$ to denote the number of independent primitive configurations of a diagram $T$, and call it the \textbf{rank} of the diagram. Similarly we can define the rank of a sub-diagram. The rank of a single vertex $v$ is $\text{deg}(v)-1$ using \textbf{DF rule 3} on page \pageref{DF rule 3}. 
The rank of a diagram equals the rank of its skeleton diagram.
In the next section, we will discuss how to determine the primitive configurations of skeleton diagrams systematically.

\section{Primitive configurations of skeleton diagrams}
\label{subsection:skeleton}

In this section, we will describe a systematic method to construct the primitive configurations of skeleton diagrams. We will choose a convenient basis for the configurations of vertices with 2 type-vv edges, and use them as basic building blocks for generic tree and loop diagrams. 
Moreover, by appropriately configuring line diagrams, the primitive configurations of a skeleton diagram can be characterized by 'sinks', such that all arrows in the reduced skeleton diagram flow toward the sink.

\subsection{Vertices with 2 type-vv edges}
A skeleton diagram can be disconnected and contains multiple components, and the primitive configurations of each components can be constructed separately. So in the following we only consider connected skeleton diagrams.

\begin{figure}
\centering
\includegraphics[scale=0.5]{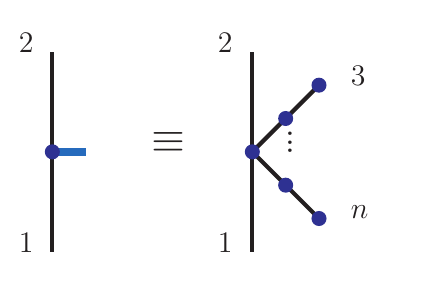}
\caption{Notation of a vertex with two type-vv edges.}
\label{fig:successiveconnected1}
\end{figure}

Let us start by considering a vertex with two type-vv edges and $n-2$ type-2v edges.
The type-vv edges will be labelled with edge-1 and edge-2, and the type-2v edges will be labelled with edge-3 to edge-$n$.
The basis of primitive configurations of the vertex can be chosen  as $\{f_{13},\ f_{23},\ f_{34},\ \cdots,\ f_{n-1,n}\}$, where $f_{ij}$ are defined in \eqref{fij-define}. We can split the basis into 3 classes labelled with 3 letters, $\texttt{a}=\{f_{13}\}$, $\texttt{b}=\{f_{23}\}$ and $\texttt{c}=\{f_{34},\ \cdots,\ f_{n-1,n}\}$.
We find it convenient to abbreviate the type-2v edges into a single bold line, as shown in Figure \ref{fig:successiveconnected1}.

\begin{figure}
\centering
\includegraphics[scale=0.5]{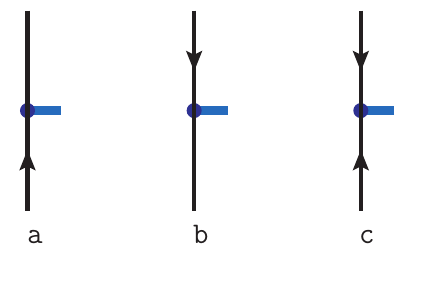}
\caption{Three types of configurations for a vertex with two type-vv edges. Notice that we have ignored the arrows carried by type-2v edges.}
\label{fig:successiveconnected2}
\end{figure}

Then the letters correspond to the three types of configurations in Figure \ref{fig:successiveconnected2}. The letter $\texttt{c}$ represents $n-3$ configurations, while each of $\texttt{a}$, $\texttt{b}$ represents a single configuration:
\begin{equation}
\mathcal{R}_{\texttt{a}}=\mathcal{R}_{\texttt{b}}=1,\ 
\mathcal{R}_{\texttt{c}}=n-3.\ 
\end{equation}
Skeleton diagrams in this abbreviated form will be called \textbf{reduced skeleton diagrams}.

\begin{figure}
\centering
\includegraphics[scale=0.6]{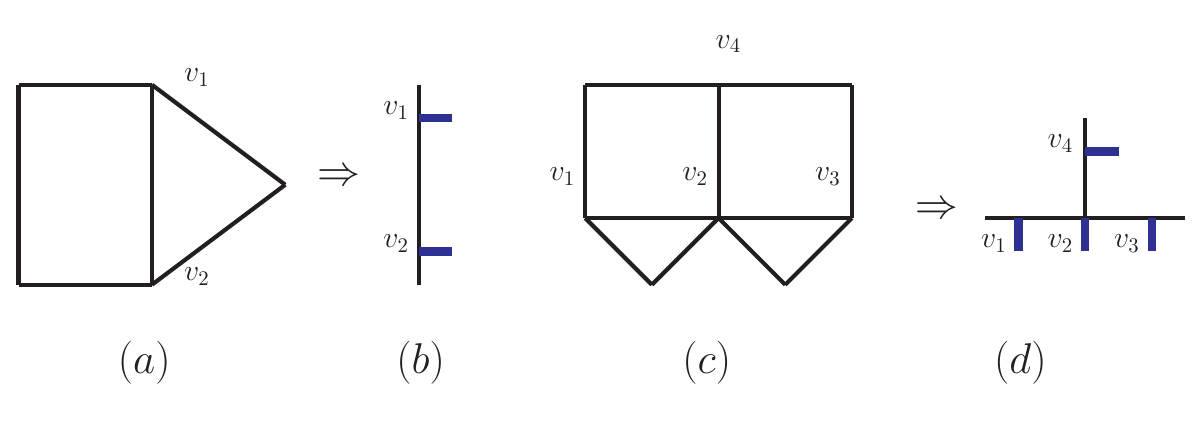}
\caption{Reduce skeleton diagrams.}
\label{fig:reduced-skeleton}
\end{figure}

The definition of reduced skeleton diagrams can also be extended to generic tree skeleton diagrams:
\begin{enumerate}
\item  If $v$ is a vertex with a exactly one type-vv edge, for example $v_1$ and $v_2$ in Figure \ref{fig:reduced-skeleton}(a),  then all but one of the type-2v edges adjacent to $v$ are replaced by a thick line. In this case, the letters $\texttt{a}, \texttt{b}, \texttt{c}$ can still be defined as discussed previous discussed. However, the type-2v edge that is not reduced now serves as one of the two type-vv edges in the earlier case.

\item If $v$ is a vertex with two or more type-vv edges, for example $v_2$ in Figure \ref{fig:reduced-skeleton}(c), then all type-2v edges adjacent to $v$ are replaced by a thick line. 

\end{enumerate}

\begin{figure}
\centering
\includegraphics[scale=0.5]{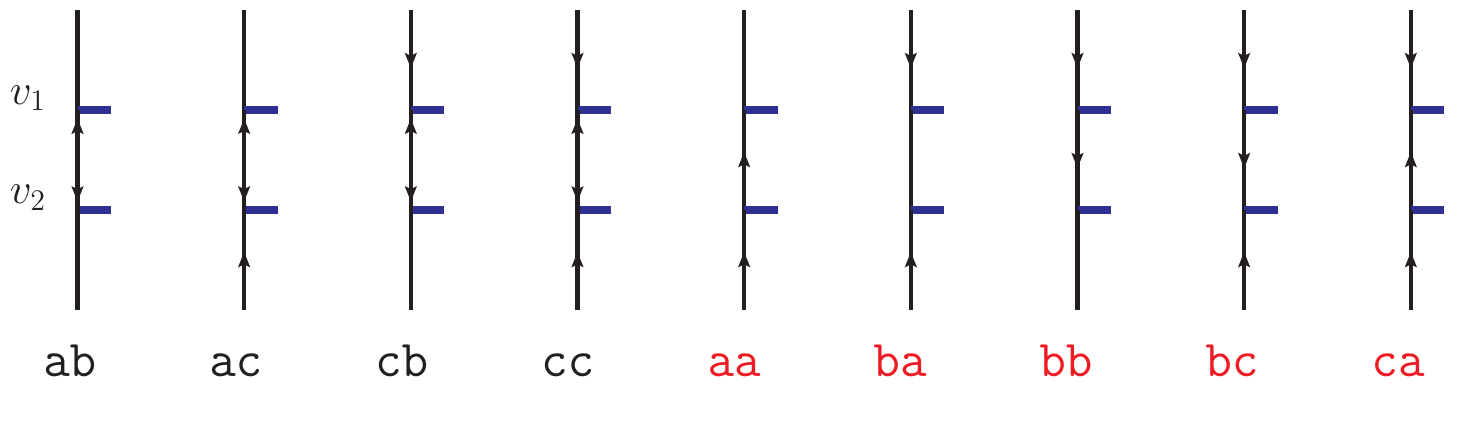}
\caption{Non-primitive and \textcolor{red}{primitive} configurations when two vertices are connnected. The first 4 configurations contain D-pairs and are thus non-primitive. The last 5 configurations (with strings in red) are primitive.}
\label{fig:join2vertex}
\end{figure}
When two vertices are connected together, as shown in Figure \ref{fig:join2vertex}, we can combine the letters \texttt{a}, \texttt{b}, \texttt{c} into strings. The strings $\mathbf{S}_1\equiv \{\texttt{ab}, \texttt{ac}, \texttt{cb}, \texttt{cc}\}$ contain $D$-pairs and are non-primitive. The other strings $\mathbf{S}_2\equiv \{\texttt{aa}, \texttt{ba}, \texttt{bb}, \texttt{bc},\texttt{ca}\}$ do not contain $D$-pairs, and in the following we will show that they form a basis of primitive configurations.

Since these 9 letters form a basis for independent configurations, $\mathbf{S}_2$ forms a basis for primitive configurations if $\mathbf{S}_1$ forms the basis for non-primitive configurations.
Let us denote the vertices by $v_1$ and $v_2$, then the rank of a string in $\mathbf{S}_1$ equals the product of ranks of the letters in the string:
\begin{equation}
\mathcal{R}\texttt{ab}=1,\ 
\mathcal{R}\texttt{ac}=\text{deg}(v_2)-3,\ 
\mathcal{R}\texttt{cb}=\text{deg}(v_1)-3,\ 
\mathcal{R}\texttt{cc}=[\text{deg}(v_1)-3][\text{deg}(v_2)-3],\ 
\end{equation}
So the total rank of these strings is $[\text{deg}(v_1)-2][\text{deg}(v_2)-2]$. In order to compute the number of independent non-primitive configurations, let us consider a new diagram which is obtained by removing the edge connecting $v_1$ and $v_2$ from the original diagram. Each non-primitive configuration of the original diagram can be obtained by a configuration of the new diagram by adding a D-pair. Since the new diagram only contains 2 vertices with degree $\text{deg}(v_1)-1$ and $\text{deg}(v_2)-1$, respectively, the total number of basis configurations is $[\text{deg}(v_1)-2][\text{deg}(v_2)-2]$. This number matches the number of non-primitive configurations in $\mathbf{S}_1$, therefore $\mathbf{S}_1$ form the basis of non-primitive configurations, and consequently the strings in $\mathbf{S}_2$ form the basis of independent primitive configurations.

We can continue to construct more complicated skeleton diagrams using the type of vertices in Figure \ref{fig:successiveconnected1}. A series of successively connected vertices form a 'line diagram', and a generic tree or loop skeleton diagram can be constructed by gluing several line diagrams together.

\subsection{Primitive configurations of  line skeleton diagram}
\label{subsubsection:line}

\begin{figure}
\centering
\includegraphics[scale=0.8, angle=0]{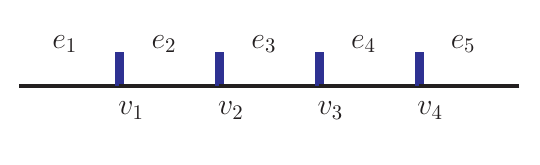}
\caption{A line diagram with 3 type-vv edges.}
\label{fig:linediagram}
\end{figure}

The simplest skeleton diagram is a line skeleton diagram formed by a series of successively connected type-vv edges. 
An example of line diagram is shown in Figure \ref{fig:linediagram}, in which $e_2$, $e_3$ and $e_4$ are type-vv edges. We have picked two  type-2v edges $e_1$ and $e_5$ on the endpoints of the line, so that we can choose and name the configurations of $v_1$ and $v_4$ following the notation in Figure \ref{fig:successiveconnected1}.

\begin{figure}
\centering
\includegraphics[scale=0.6]{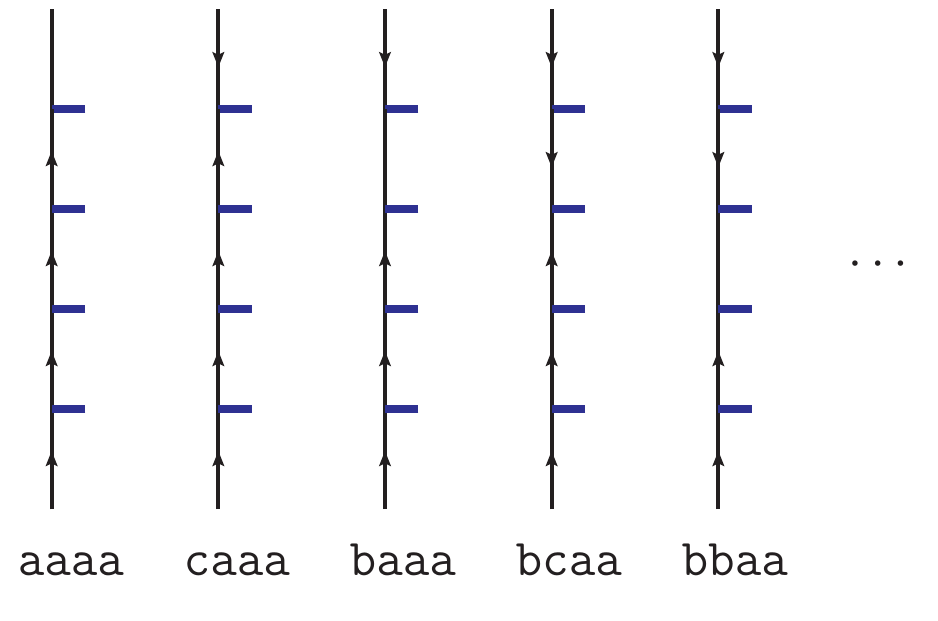}
\caption{A subset of primitive configurations of a line with 4 vertices.}
\label{fig:join4vertex}
\end{figure}

Some primitive configurations of Figure \ref{fig:linediagram} are shown in Figure \ref{fig:join4vertex}.

For a configuration to be primitive, the string must satisfy specific constraints: every occurrence of \texttt{a} or \texttt{c} must be immediately followed by \texttt{a}, and every occurrence of \texttt{b} or \texttt{c} must be immediately preceded by \texttt{b}. Additionally, the directional patterns of the arrows must converge such that all arrows flow into a designated region of the diagram.
For \texttt{caaa} and \texttt{bcaa} in Figure \ref{fig:join4vertex}, all arrows flow into a vertex.
For \texttt{baaa} and \texttt{bbaa}, all arrows flow into an undirected edge.

This feature can be extended to generic skeleton diagrams. Primitive configurations can be choosen so that all arrows uniformly converge toward specific regions which can be vertices, edges or loops. These regions will be called \textbf{sinks}.

As can be seen from Figure \ref{fig:join4vertex}, all edges in the reduced skeleton diagram are directed except for the sink edges. As will be discussed later, this is true for a generic reduced skeleton diagram, as long as each vertex of the diagram has at least one external legs. The original skeleton diagram does contain more undirected exernal legs, but these legs are not exhibited in the reduced skeleton diagram.

If the sink is a vertex $v_i$, then this vertex will be type-\texttt{c}, and there are $\text{deg}(v_i)-3$ configurations.
If the sink is an edge, the string only contain \texttt{a} and \texttt{b}, and there is only one configuration. There are in all $V+1$ edges in the line-diagram, so the total number of configurations is
\begin{equation}\label{rank-line-1}
\mathcal{R}_{\text{line}}=\sum_{i=1}^V[\text{deg}(v_i)-3]+V+1\ .
\end{equation}
We will use $\mathcal{E}_{\text{line}}$ to denote the number of external legs of the line, then using the relation between edges and degree of vertices,
\begin{equation}\label{rank-line-2}
\sum_{i=1}^V\text{deg}(v_i)=2(V-1)+\mathcal{E}_{\text{line}}.\ 
\end{equation}
Combining \eqref{rank-line-1} and \eqref{rank-line-2} we derive
\begin{equation}\label{rank-line-3}
\mathcal{R}_{\text{line}}=\mathcal{E}_{\text{line}}-1\ .
\end{equation}

As an example, the skeleton diagram in Figure \ref{fig:join2vertex}(b) is a line diagram with two vertices of degree 3. The reduced skeleton diagram contains 3 edges; each can be selected as a sink, and each corresponds to one configuration. If a vertex is chosen as the sink, it corresponds to $\text{deg}(v_i)-3=0$ configurations. This diagram thus has three configurations, which is also consistent with \eqref{rank-line-3}, given that the diagram has 4 external legs.

We observe that \textbf{DF rule 3} can also be regarded as a special case of \eqref{rank-line-3}.
In fact, as will be seen in the next subsection, \eqref{rank-line-3} also hold for generic tree skeleton diagrams.

\subsection{Primitive configurations of  tree skeleton diagrams}
\label{subsubsection:tree}

\begin{figure}
\centering
\includegraphics[scale=0.6]{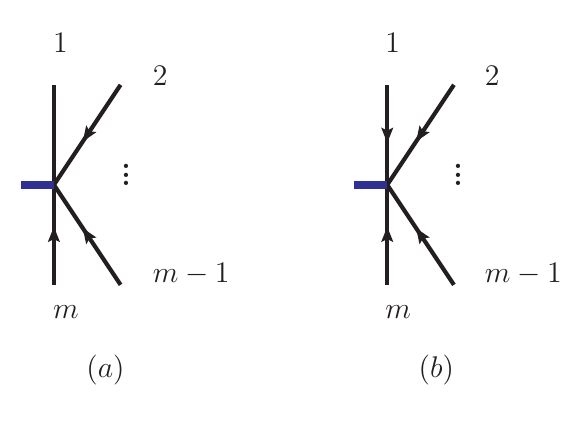}
\caption{$m$ lines attached to the same degree-$n$ vertex. The edges $(m+1)$ to $n$ are type-2v edges, and are represented by a thick line. The arrows only show the configurations of the center vertex.}
\label{fig:centervertex}
\end{figure}

Skeleton tree diagrams can be constructed by gluing several skeleton line diagrams together.
Let us start with $m$ lines labelled with line-1 to line-$m$, and glue them to a degree-$n$ vertex with $n>m$, as shown in Figure \ref{fig:centervertex}.

The degree-$n$ vertex will be called the center-vertex. We choose the configurations of the center-vertex to be $\{f_{1,m+1},\ f_{2,m+1},\cdots,f_{m,m+1},\ f_{m+1,m+2},\cdots,\ f_{n-1,n}\}$. The configuration $f_{1,m+1}$ corresponds to Figure \ref{fig:centervertex}(a), while the configuration $f_{i,i+1}$ with $m+1\le i\le n-1$, correspond to Figure \ref{fig:centervertex}(b).

Not all configurations of a line are allowed in the tree diagram, because they must be consistent with the configuration of the center-vertex. For example, consider line-2 in Figure \ref{fig:centervertex}(a). The arrow on the edge connecting line-2 and the center-vertex implies that all arrows of line-2 must flow towards the center vertex (otherwise the edge becomes a D-pair), so the configuration of line-2 is completely fixed. The same happens for line-3 to line-$m$. On the other hand, there is no extra constraint imposed on the configuration of line-1, so all configurations of line-1 are allowed. There are $\mathcal{E}_{\text{line-1}}-1$ such configurations, and the sinks of line-1 are also the sinks of the full diagram. Similarly,  for other $f_{i,m+1}$ with $i\le m$, the sinks of the diagram is the sink of line-$i$. 

For the $f_{i,i+1}$ configuration in Figure \ref{fig:centervertex}(b), all arrows on the lines are fixed to be incoming. There are $n-m-1$ such configurations and the sink is the center-vertex.

The total rank of the diagram is 
\begin{equation}
\mathcal{R}_{\text{tree}}=n-m-1+\sum_{i=1}^m(\mathcal{E}_{\text{line-}i}-1)\ .
\end{equation}
Notice that there are $n-m$ external legs attached to the center-vertex, and the external legs of line-$i$ are also the external legs of the tree diagram, except the one connects the line and the vertex. So the diagram has in all $\mathcal{E}_{\text{tree}}=n-m+\sum_{i=1}^m(\mathcal{E}_{\text{line-}i}-1)$ exernal legs, and the rank still satisfies:
\begin{equation}\label{rank-tree-1}
\mathcal{R}_{\text{tree}}=\mathcal{E}_{\text{tree}}-1\ .
\end{equation}

\begin{figure}
\centering
\includegraphics[scale=0.45]{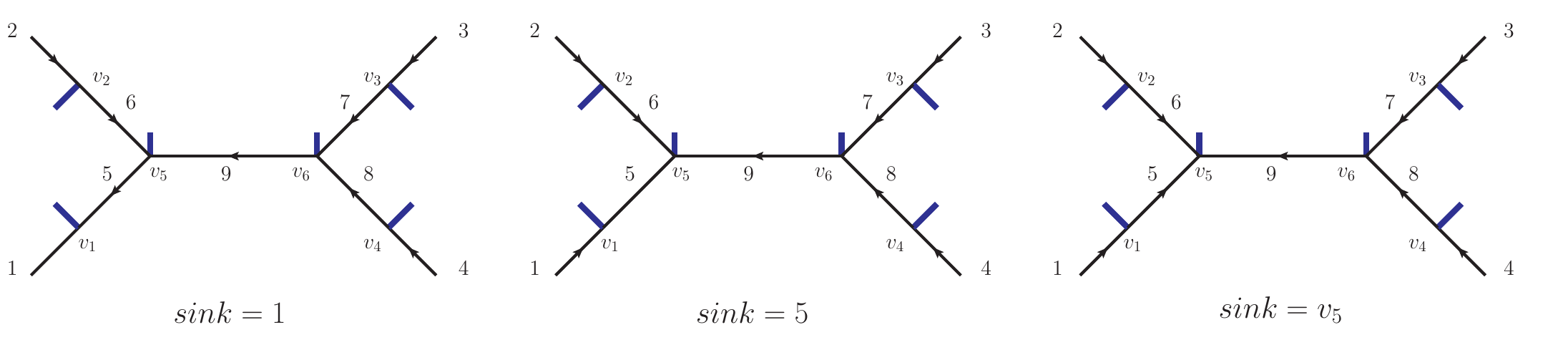}
\caption{The sinks and configurations of a tree diagram.}
\label{fig:tree-config}
\end{figure}
The configuration of more general tree diagrams can be constructed similarly using the sink technique. As an example, in Figure \ref{fig:tree-config}, all vertices and edges in the reduced skeleton diagram can be sinks. The configurations in the Figure correspond to $\text{sink}=(\text{edge-1},\text{edge-5}, v_5)$ respectively.
It can be checked that \eqref{rank-tree-1} still holds for generic tree diagrams.

\begin{figure}
\centering
\includegraphics[scale=0.6]{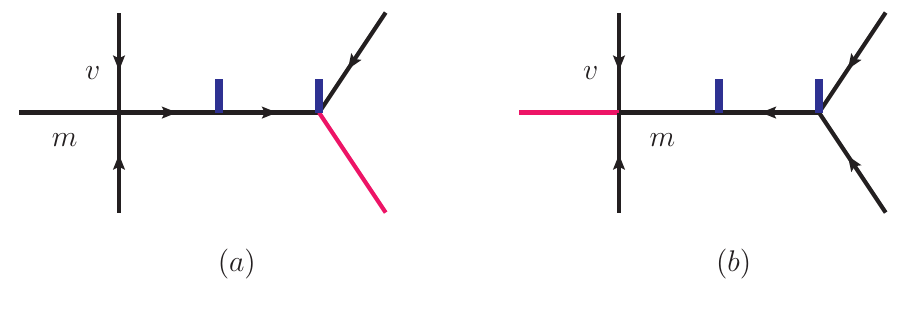}
\caption{The sinks in the presence of a saturated vertex. The red lines are sinks. The edges labelled $m$ are not sinks, but at the same time have no arrows. }
\label{fig:saturatedconverge}
\end{figure}

Special attention should be paid to the case where $m$ lines are attached to a a central vertex of degree $n=m$; this vertex will be referred to as a \textbf{saturated vertex}. The key distinction here is that no type-2v edge adjacent to the central vertex in this scenario. Consequently, among the $m$ type-vv edges adjacent to the center vertex, two must be arrow-free. The configurations of the central vertex discussed earlier, which assume $n>m$, can no longer be applied here.
For example, the vertex $v$ in Figure \ref{fig:saturatedconverge} is a saturated vertex, which has no thick blue edge attached.
We can first pretend that line-$m$ is a type-2v edge, and find the configurations of the other lines following the method above; then we determine all allowed configurations of line-$m$.
We will not go into the details, but only present the rules to determine sinks in the presence of saturated vertices:
\begin{enumerate}

\item If a saturated vertex is the sink, then from our previous discussion, it corresponds to $n-m-1\rightarrow -1$ configuration when $n\rightarrow m$. So the saturated vertex cannot be the sink.

\item Among the edges adjacent to a saturated vertex, exactly one must remain arrow-free\footnote{Any such edge may be freely selected to be arrow-free, as all choices yield equivalent results by virtue of the Bianchi identity.}. Notably, this arrow-free edge cannot be considered as part of the sink. For example, in Figure \ref{fig:saturatedconverge}(a), edge-$m$ is arrow-free, but not all arrows flow towards it.

\item If an edge adjacent to the saturated vertex is designated as the sink (as illustrated in Figure \ref{fig:saturatedconverge}(b)), we must select another adjacent edge to remain arrow-free. This results in a pair of arrow-free edges adjacent to the saturated vertex. Suppose there are $m$ edges adjacent to the saturated vertex. The total number of ways to choose two distinct edges as arrow-free is $\frac{1}{2}m(m-1)$. However, the Bianchi identity imposes constraints that render only $(m-1)$ of these configurations independent. Consequently, for simplicity, we may fix the pair $(i,m)$ with $i=1,\cdots, m-1$ as the designated arrow-free edges.

\end{enumerate}

All arrows on the other edges still flow to the sink, and \eqref{rank-tree-1} still holds in the presence of saturated vertices. Diagrams with saturated vertices start to appear at length-7.


\subsection{Primitive configurations of loop skeleton diagrams}
\label{subsubsection:loop}

\begin{figure}
\centering
\includegraphics[scale=0.7]{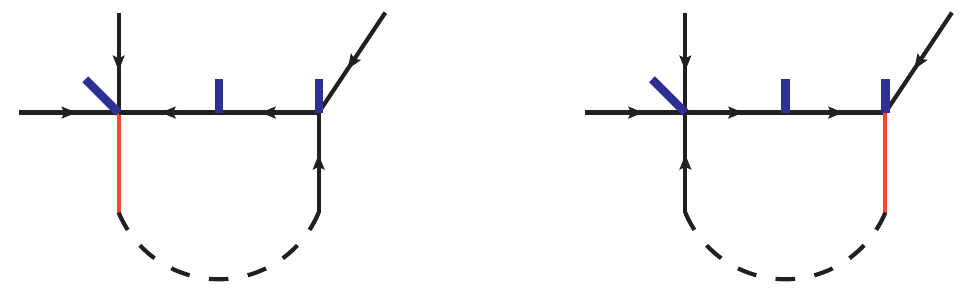}
\caption{Glue a tree into a loop. The red edges are the sinks of the original tree.}
\label{fig:gluetree}
\end{figure}

One-loop skeleton diagrams can be obtained by gluing two external legs of tree skeleton diagrams. In order for a primitive configuration of the original tree skeleton diagram to remain primitive after the gluing, at least one of these two external legs shall be arrow-free. However, this condition cannot be satisfied by the majority of configurations, since the arrows are pointing towards the sinks. Actually, as long as the digram has no saturated vertices, an external leg carries no arrow only if it is the sink itself. So there are only two configurations for a one loop diagram, as illustrated in Figure \ref{fig:gluetree}.

\begin{figure}
\centering
\includegraphics[scale=0.7]{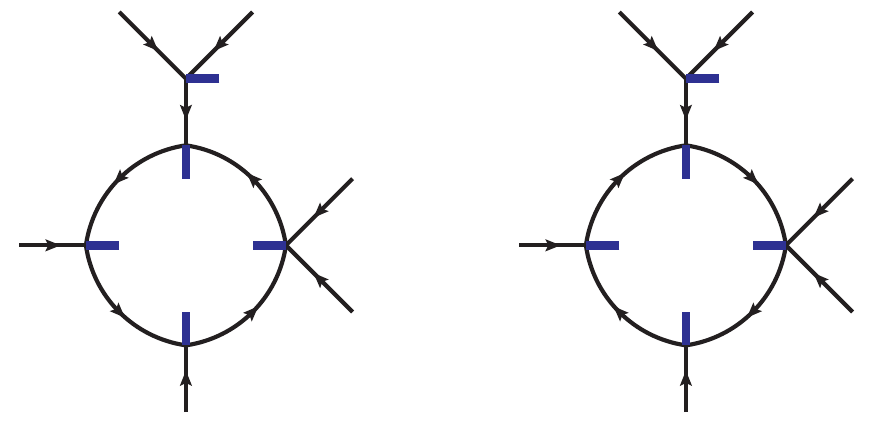}
\caption{Whirlpools: the primitive configurations of 1-loop skeleton diagram.}
\label{fig:whirlpools}
\end{figure}

Therefore, a one-loop skeleton diagram has at most two primitive configurations which have the shape of whirlpools, as shown in Figure \ref{fig:whirlpools}. 
Arrows not on the loop point towards the loop, while arrows on the loop are oriented in a clockwise or counter-clockwise direction.

One may try to glue two external legs of a 1-loop skeleton diagram to form a 2-loop skeleton diagram. However, there is no primitive configurations since all external edges of the 1-loop skeleton diagram carry arrows.
For the same reason, there are no primitive configurations for any higher loop skeleton diagram.

\begin{figure}
\centering
\includegraphics[scale=0.7]{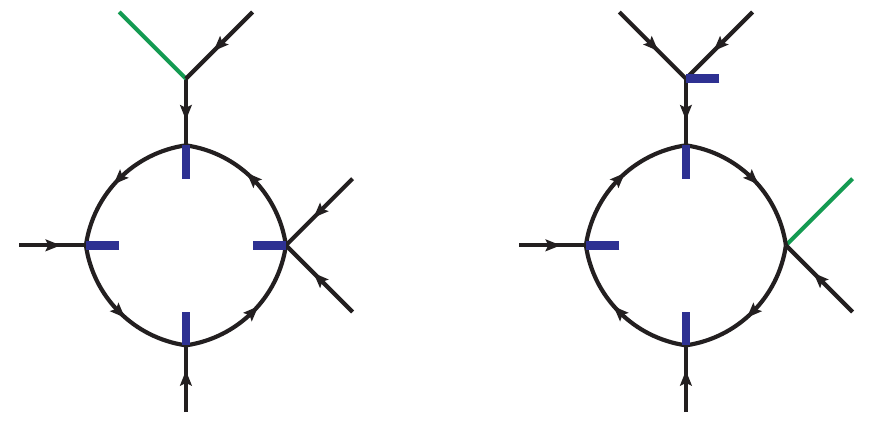}
\caption{1-loop primitive configurations in presence of saturated vertices. The arrow on one of the edges (colored green) adjacent to the saturated vertice shall be removed.}
\label{fig:whirlpool-sat}
\end{figure}

In  the presence of saturated vertices, the structure of sinks of loop skeleton diagrams are still the same as Figure \ref{fig:whirlpools}, except that there can be more undirected edges similar as the tree case, as shown in Figure \ref{fig:whirlpool-sat}.

\begin{figure}
\centering
\includegraphics[scale=0.6]{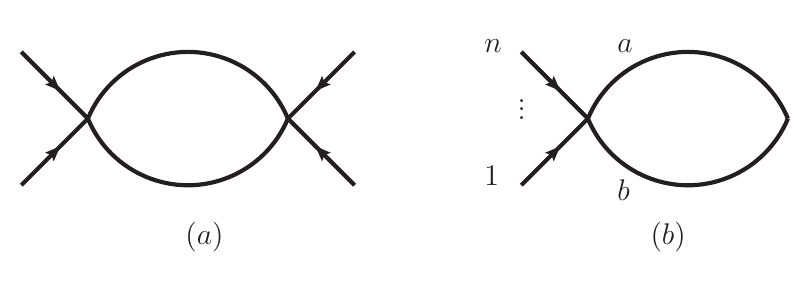}
\caption{Some special cases of one-loop diagrams.}
\label{fig:special1loop}
\end{figure}

Now let us consider the special case when the diagram contains bubbles, as shown in Figure \ref{fig:special1loop}. As discussed in Section \ref{subsection:op-diagram}, the diagram is invariant under the permutation of the bubble edges, which imposes more constraints on the primitive configurations.

In Figure \ref{fig:special1loop}(a), the clockwise or counter-clockwise configurations are equivalent since they are related by a permutation of two loop edges. It is more convenient to move the arrows in the loop to the external legs, for example:
\begin{equation}
D_1F_{24}D_2F_{13}
=D_1F_{24}D_1F_{23}+D_1F_{24}D_3F_{12}\sim D_1F_{24}D_3F_{12}
=-\frac{1}{2}D_4F_{12}D_3F_{12}
\end{equation}

Figure \ref{fig:special1loop}(b) contains a skeleton snail which was discussed in Figure \ref{fig:np-loop}. 
It can be recoganized as a special case of Figure \ref{fig:special1loop}(a), and the only independent primitive configuration of this structure is $D_{1\cdots n}F_{ab}F_{ab}$.




\section{Construction of primitive operator basis}
\label{section:primitive-basis}

\begin{figure}
\centering
\includegraphics[scale=0.6]{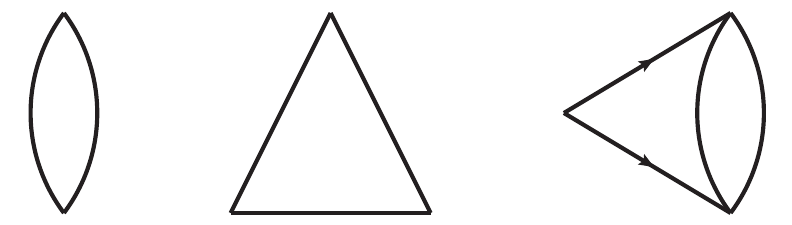}
\caption{Length-2 and length-3 primitive operator diagrams.}
\label{fig:length-2-3-repeat}
\end{figure}

With the help of operator diagrams, the complete list of length-$n$ primitive operators can be obtained by first finding all undirected diagrams with $n$ vertices and $\mathcal{R}(T)\ge 1$, and then generating the primitive configurations of each diagram by examining its skeleton diagram.
The length-2,3 operator diagrams are shown in Figure \ref{fig:length-2-3-repeat}.

In this section we first present the primitive diagrams with length-4 and length-5, and discuss their symmetries. In Section \ref{subsection:len<=7} we give the number of primitive operators with length $\le 7$. Last, in Section \ref{subsection:refine}, we show that the primitive configurations can be further refined according to the symmetries of diagrams, which makes it easier to construct the complete operator basis when the vertex permutations of the diagrams are taken into account.

\subsection{Length-4 primitive operators}
\label{subsection:higher-length-4}
The length-$n$ diagrams satisfying $\mathcal{R}(T)\ge1$ can be systematically constructed from skeleton diagrams with at most $n$ vertices, where skeleton diagrams can be derived from vacuum diagrams without degree-2 vertices. A detailed algorithm for this procedure is provided in Appendix \ref{appendix:construct}.

\begin{figure}
\centering
\includegraphics[scale=0.5]{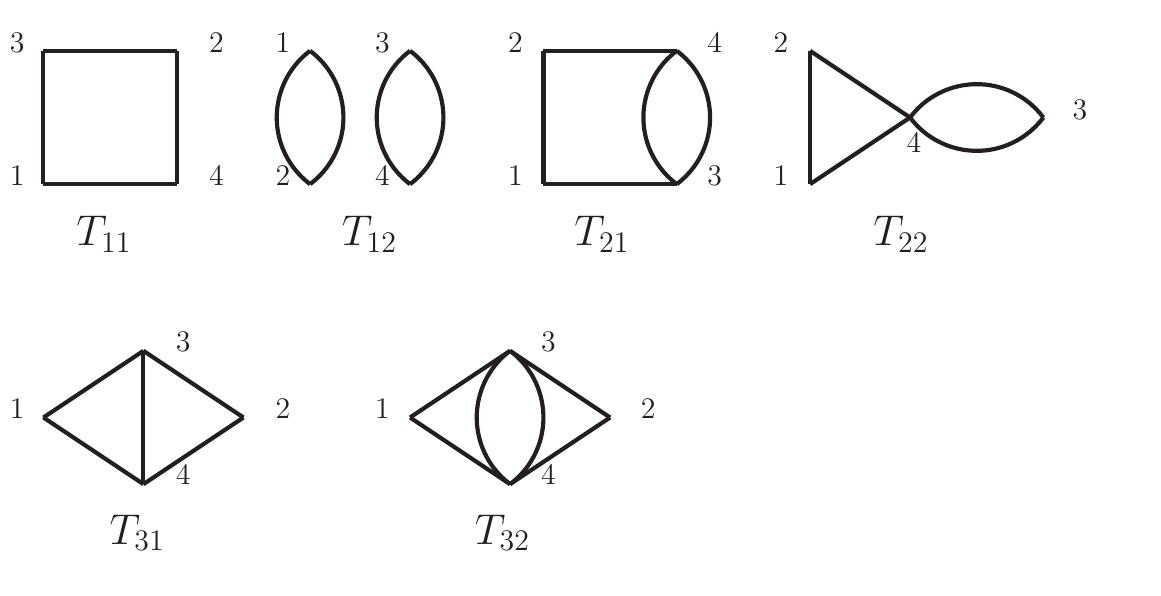}
\caption{The length-4 operator diagrams containing primitive configurations.}
\label{fig:len4topo}
\end{figure}

Following the algorithm, we obtained the length-4 undirected diagrams with nonzero rank, which are given in Figure \ref{fig:len4topo}.

The diagrams are labelled in the form of $T_{xy}$. $x$ is the type of symmetry of the diagram which will be discussed further in Section \ref{subsection:refine}, and $y$ is a label to discriminate diagrams which have the same symmetry. For example, $T_{11}$ and $T_{12}$ are two diagrams with the same symmetry. 
The ranks and symmetries of the diagrams are listed in Table \ref{symmetry-group-length-4}. 
\begin{table}
\centering
\begin{tabular}{|c |c|c|c|c|c|c|}
\hline
$T$ & $T_{11}$ &$T_{12}$ &$T_{21}$&$T_{22}$&$T_{31}$&$T_{32}$   \\
\hline
$\mathcal{R}$ & 1&1&1&1&3 &1  \\
\hline
$G_T$ & $D_4$&$D_4$&$S_2$&$S_2$&$S_2\times S_2$ &$S_2\times S_2$  \\
\hline
$|G_T|$ & 8&8&2&2&4&4  \\
\hline
\end{tabular}
\caption{The  ranks and symmetry groups of length-4 operator diagrams.}
\label{symmetry-group-length-4}
\end{table}

\begin{figure}
\centering
\includegraphics[scale=0.5]{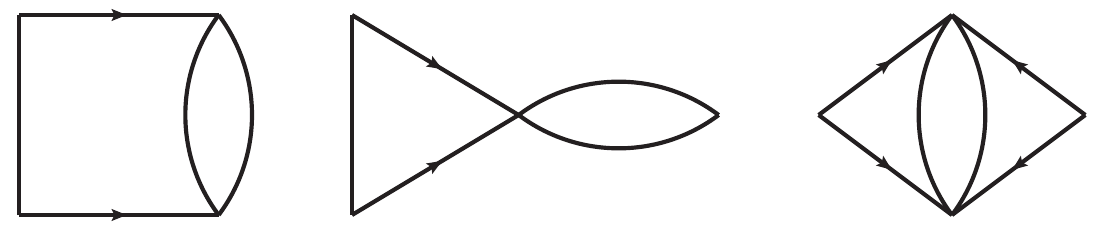}
\caption{The primitive configurations of $T_{21}$, $T_{22}$ and $T_{32}$.}
\label{fig:primitive-4-bubble}
\end{figure}

In what follow, we discuss the configurations of these diagrams in detail. The configurations of $T_{11}$ and $T_{12}$ are arrow-free. For $T_{21}$, $T_{22}$ and $T_{32}$, each of which contains a 1-loop bubble structure; accordingly, as discussed in Section \ref{subsubsection:loop},  each possesses only one primitive configuration, as illustrated in Figure \ref{fig:primitive-4-bubble}. 

\begin{figure}
\centering
\includegraphics[scale=0.5]{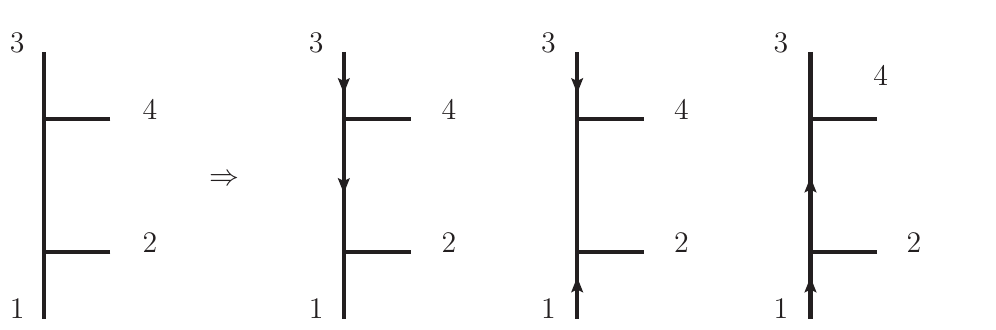}
\caption{The primitive configurations of the skeleton diagram of $T_{31}$.}
\label{fig:T41fromsubdiagram}
\end{figure}
\begin{figure}
\centering
\includegraphics[scale=0.5]{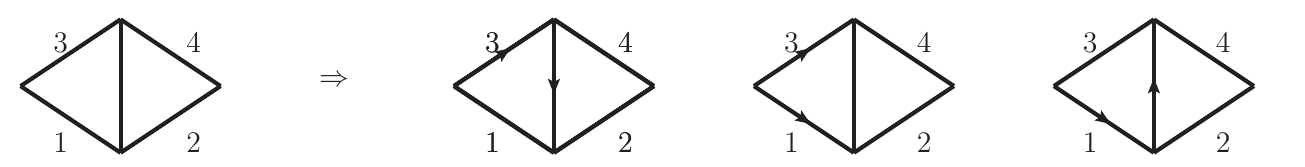}
\caption{The primitive configurations from the skeleton diagram approach.}
\label{fig:operatordiagramT41}
\end{figure}

$T_{31}$ is the only length-4 topology which requires additional inspection.
The skeleton diagram is shown in Figure \ref{fig:T41fromsubdiagram}, together with the 3 primitive configurations obtained using the sink technique.\footnote{All sinks of this diagram are edges. The primitive configurations whose sinks are vertices do not exist because $v-3=0$.} We have labelled the external edges of the sub-diagram by numbers, in order to show how the sub-diagram is embeded in the whole diagram, as shown in Figure \ref{fig:operatordiagramT41}. The topology also has a non-primitive configuration, in which the edge in the center is a $D$-pair.

\subsection{Length-5 primitive operators}
\label{subsection:higher-length-5}

\begin{figure}
\centering
\includegraphics[scale=0.5]{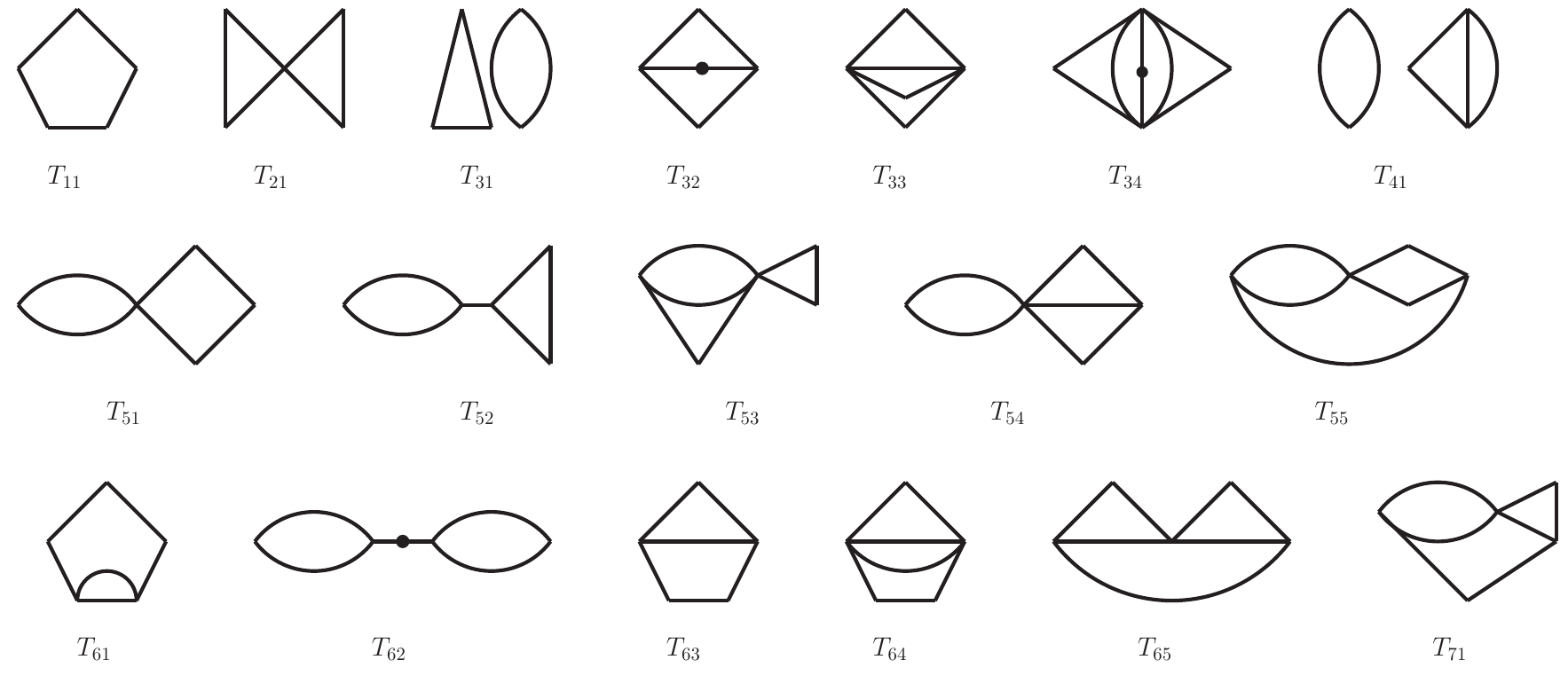}
\caption{Length-5 undirect diagrams with primitive configurations.}
\label{fig:len5op1}
\end{figure}

Compared with the length-4 case, at length-5 there are far more operators and richer structures.
The connected diagrams with $\mathcal{R}\ge 1$ are shown in Figure \ref{fig:len5op1}, and there are 16 connected diagrams and 2 non-connected diagram. 

The diagrams are again classified by their symmetry groups, shown in Table \ref{symmetry-group-length-5}.

\medskip{}
\begin{table}
\centering
\begin{tabular}{|c |c|c|c|c|c|c|c|}
\hline
$T$ & $T_{1}$ &$T_{2}$ &$T_{3}$&$T_{4}$&$T_{5}$&$T_{6}$ &$T_{7}$   \\
\hline
$\mathcal{R}$ &1 &3&(1,4,5,1)&1&(1,1,1,1,1)&(1,1,3,1,2) &1   \\
\hline
$G_T$ & $D_5$&$D_4\times S_1$&$S_3\times S_2$&$S_1\times S_2\times S_2$&$S_1^3\times S_2$ &$ \{1,(12)(34)\}\times S_1$&$S_1^5$  \\
\hline
$|G_T|$ & 10&8&12&4&2&2&1 \\
\hline
\end{tabular}
\caption{The ranks and symmetry groups of length-5 operator diagrams. (1,4,5,1) means $\mathcal{R}(T_{31})=1,\ \mathcal{R}(T_{32})=4,\ \cdots$.}
\label{symmetry-group-length-5}
\end{table}
\medskip{}

The primitive configurations for each diagram in DF-form are presented in Table \ref{len-5-prim} in Appendix \ref{appendix:len-5}. For instance, the skeleton diagram of $T_{21}$ consists of a single vertex with degree 4, thus the number of pritmitive configurations, denoted as $\mathcal{R}(T_{21})$, is 3. The primitive configurations can be chosen as
\begin{equation}
\begin{aligned}
P_{21}^1=&F_{12}F_{23}F_{45}F_{56}D_{13}F_{46},\\
(13)(24)P_{21}^1=&F_{45}F_{56}F_{12}F_{23}D_{13}F_{46},\\
P_{21}^2= &\{F_{12},F_{23}\}\{F_{45},F_{56}\}D_{14}F_{36}\ .\\
\end{aligned}
\end{equation}

We have also constructed all length$\le 7$ undirected diagrams with $\mathcal{R}\ge1$. 
The number of connected diagrams with $\mathcal{R}\ge1$ are

\medskip{}
\begin{tabular}{|c |c|c|c|c|c|c|}
\hline
Length & 2 &3&4&5&6&7  \\
\hline
$\#$ connected & 1 &2&5&16&58&226  \\
\hline
\end{tabular}.
\medskip{}

\subsection{The total number of primitive operators with length $\le 7$}
\label{subsection:len<=7}

The total number of primitive operators of length-$n$ can be determined by summing the ranks of all primitive diagrams of length-$n$.
As discussed in the previous section, two operator diagrams related by a vertex permutation should be treated as distinct diagrams, unless the diagram is invariant under that permutation ( i.e., the permutation is an element of $G_T$). Starting from an undirected operator diagram $T$ with $n$ vertices, we can thus generate $\frac{n!}{|G_T|}$ distinct diagrams, denoted by $S_nT$. 

Suppose the diagram $T$ has $\mathcal{R}_T$ primitive configurations, then the total number of primitive configurations in $S_nT$ is
\begin{equation}
\mathcal{R}(S_nT)=\frac{n!}{|G_T|}\mathcal{R}_T\ .
\end{equation}

Let $\mathbf{C}_n$ be the set of connected length-$n$ diagrams with $\mathcal{R}\ge 1$, then the total number of primitive configurations in these diagrams are
\begin{equation}
\mathcal{R}(S_n\mathbf{C}_n)=n!\sum_{T\in \mathbf{C}_n}\frac{1}{|G_T|}\mathcal{R}_T\ .
\end{equation}
The values of $\mathcal{R}(S_n\mathbf{C}_n)$ for different $n$ are as follows:

\medskip{}
\begin{tabular}{|c |c|c|c|c|c|c|}
\hline
$n$ & 2 &3&4&5&6&7  \\
\hline
$\mathcal{R}(S_n\mathbf{C}_n)$ & 1 &4&51&1057&27765&925830  \\
\hline
\end{tabular}.
\medskip{}

As an example, let us consider the case of $n=4$. The connected diagrams are presented in Figure \ref{fig:len4topo}. However, the second diagram $T_{21}$ is a disconnected diagram and thus neglected here.
The ranks and symmetry groups of these connected diagrams are provided in Table \ref{symmetry-group-length-4}. Using this information, we can calculate the total number of primitive configurations, which is given by:
\begin{equation}
4! \left( \frac{1}{8} + \frac{1}{2} + \frac{1}{2} + \frac{3}{4} + \frac{1}{4} \right) = 51.\end{equation}

Some detailed discussions on length-5 primitive configurations  can be found in Appendix \ref{appendix:len-5}.

The rank of disconnected diagrams can be computed from $\mathcal{R}(S_n\mathbf{C}_n)$.
Let $\mathbf{D}=\mathbf{C}_{n_1}\cup\cdots \cup \mathbf{C}_{n_k}$ be the set of disconnected diagrams with $k$ components with length $(n_1,\cdots, n_k)$, the rank of $\mathbf{D}$ is
\begin{equation}
\mathcal{R}(\mathbf{D})
=\frac{(n_1+\cdots +n_k)!}{n_1!\cdots n_k!S_{n_1\cdots n_k}}
\mathcal{R}(S_{n_1}\mathbf{C}_{n_1})\cdots \mathcal{R}(S_{n_k}\mathbf{C}_{n_k})
\end{equation}
in which $S_{n_1\cdots n_k}$ is the symmetry factor of $(n_1,\cdots, n_k)$.

 For example, for diagrams with two length-2 and one length-3 subdiagrams, the total number of primitive configurations is
\begin{equation}
\frac{1}{2}\frac{7!}{2!2!3!}\mathcal{R}^2(S_2\mathbf{C}_2)\mathcal{R}(S_3\mathbf{C}_3)=420\ ,
\end{equation}
in which $\frac{1}{2}$ is from the symmetry factor of $(2,2,3)$.

Including the contribution of disconnected diagrams, the total number of length-$n$ primitive operators are given by:

\medskip{}
\begin{tabular}{|c |c|c|c|c|c|c|}
\hline
$n$ & 2 &3&4&5&6&7  \\
\hline
$\mathcal{R}_n$ & 1 &4&54&1097&28705&955587  \\
\hline
\end{tabular}\ .
\medskip{}

 The study of primitive operators can also be approached from an algebraic perspective. Kinematic form factors are viewed as polynomial vectors, with primitive form factors corresponding to the generators of the kinematic form factor module. These primitive form factors can be derived via straightforward algebraic algorithms. Using such algorithms, we constructed primitive form factors ranging from length-2 to length-5 and observed a perfect match between the operator counts obtained through algebraic and diagrammatic methods. Further details on the algebraic framework are provided in Appendix \ref{algebraic}.

\subsection{Refine primitive operators using the symmetry of the diagram}
\label{subsection:refine}



In previous sections, we established an algorithm for constructing arrow configurations via the skeleton diagram method. Here, we refine this approach by systematically arranging primitive configurations as representations of the symmetry group associated with the diagram. This reorganization not only simplifies the operator but also improves the efficiency of subsequent procedures, such as dressing color factors.

We continue to use the length-4 operators as illustrative examples. In Section \ref{subsection:higher-length-4}, the configurations of these length-4 operators are constructed using the sink technique. However, these primitive configurations, such as those depicted in Figure \ref{fig:T41fromsubdiagram}, generally fail to preserve the symmetry of the diagram. 
The action of $G_T=S_2\times S_2$ on these configurations yields 6 distinct operators, among which 4 are linearly independent. Given that there are only 3 primitive configurations, this implies that a linear combination of these 6 operators must be non-primitive.
Thus, it is more desirable to identify a set of primitive operators that remains invariant under the action of $G_T$. 
More precisely, let us denote the linear space spanned by the operators produced by acting $G_T$ on $P$ as $\text{span}(G_TP)$. We would like to find a list of primitive operators $P_1,\cdots, P_N$, so that the linear space spanned by independent primitive operators can be decomposed into the direct sum of $\text{span}(G_TP_N)$,
\begin{equation}
L_{p}^T=\text{span}(G_TP_1)\oplus\cdots\oplus\text{span}(G_TP_N)\ .
\end{equation}

We start with the operators obtained using the skeleton digram approach. From them we choose a candidate primitive operator $P_1'$,
and define $W_1=\text{span}(G_TP_1')\cap L_{np}^T$, which is the non-primitive component in $\text{span}(G_TP_1')$. If $W_1\ne\{0\}$, we should remove $W_1$ from $\text{span}(G_TP_1')$ by  decomposing $\text{span}(G_TP_1')=W_1\oplus \text{span}(G_TP_1)$. The decomposition can be achieve using the the Maschke's theorem (see Appendix \ref{Maschke}). Thus we have determined the first preferred primitive operator $P_1$.

If $\text{dim}\Bigl[\text{span}(G_TP_1)]< \text{dim}\Bigl[L_{p}^T\Bigr]$, we choose another candidate primitive operators $P_2'\notin L_{np}^T\oplus \text{span}(G_TP_1)$, and define $W_2=\text{span}(G_TP_2')\cap \Bigl[L_{np}^T\oplus \text{span}(G_TP_1)\Bigr]$. If $W_2\ne\{0\}$, we decompose $\text{span}(G_TP_2')=W_2\oplus \text{span}(G_TP_2)$, and determine the second preferred primitive operator $P_2$. Similarly we can find more preferred primitive operators, until a complete basis of $L_{p}^T$ is obtained.

\begin{figure}
\centering
\includegraphics[scale=0.5]{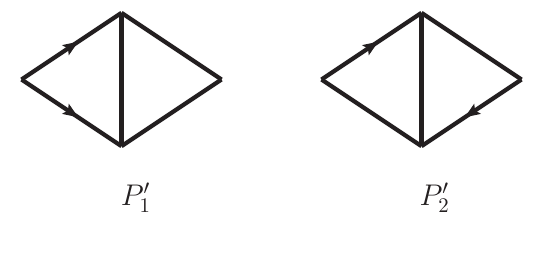}
\caption{Condidate primitive operators for $T_{41}$.}
\label{fig:arrowconfigurations}
\end{figure}
\begin{figure}
\centering
\includegraphics[scale=0.5]{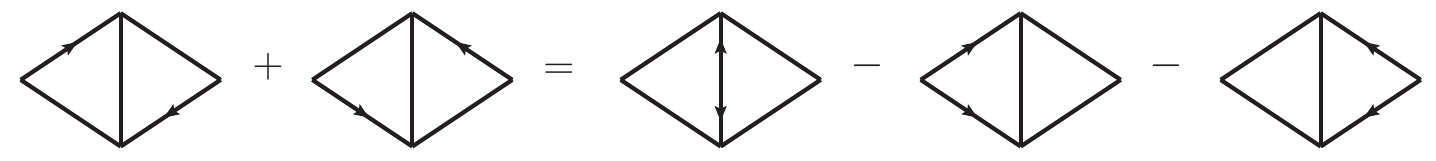}
\caption{A relation in $T_{31}$ configurations.}
\label{fig:T41relations}
\end{figure}

Let us illustrate this method by applying it to the topology $T_{31}$. We choose the candidate operators $P_1'$ and $P_2'$ in Figure \ref{fig:arrowconfigurations}. $G_TP_1'=\{1,\ (12)\}P_1'$ contains two elements and $W_1=\{0\}$, therefore $P_1=P_1'$.
$G_TP_2'=\{1,\ (12)\}P_2'$ also contains two elements, but $W_2\ne \{0\}$ because these configurations satisfy a relation shown in Figure \ref{fig:T41relations}. We should replace $P_2'$ by $P_2=P_2'-(12)P_2'$. 

\begin{figure}
\centering
\includegraphics[scale=0.5]{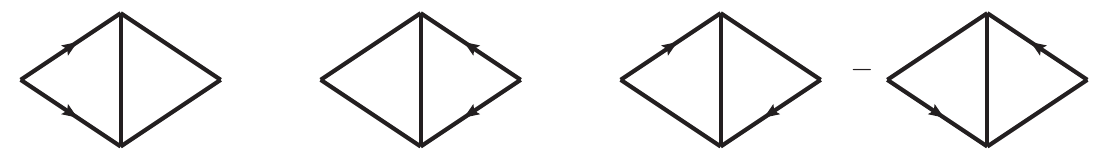}
\caption{A set of primitive operators which is invariant under the action of $S_2\times S_2$.}
\label{fig:T41preferprimitive}
\end{figure}

Since $T_{31}$ contains three primitive configurations, these operators forms a complete set of preferred primitive operators, as illustrated in Figure \ref{fig:T41preferprimitive}.


For completeness, here we also list the DF-block form of length-4 primitive operator which are refined by the symmetry of operator diagrams:
\begin{equation}
\begin{aligned}
&P_{11}^1=F_{12}F_{34}F_{23}F_{41}\ ,\\
&P_{12}^1=F_{12}F_{12}F_{34}F_{34}\ ,\\
&P_{21}^1=F_{12}F_{23}D_1F_{45}D_3F_{45}\ ,\\
&P_{22}^1=F_{12}F_{23}F_{45}D_{13}F_{45}\ ,\\
&P_{31}^1=F_{12}F_{34}D_1F_{35}D_{2}F_{45}\ ,\\
&P_{31}^2=F_{34}F_{12}D_1F_{35}D_{2}F_{45}\ ,\\
&P_{31}^3=[F_{12},F_{34}]D_1F_{35}D_{4}F_{25}\ ,\\
&P_{32}^1=F_{12}F_{34}D_{13}F_{56}D_{24}F_{56}\ ,\\
\end{aligned}\label{prim4list}
\end{equation}
in which $P^i_{xy}$ corresponds to the $i$-th primitive configuration of the diagram $T_{xy}$.

Following the same steps, we have also explicitly constructed all length$\le 6$ primitive operators which respect $G_T$ symmetry.

\section{Discussion}
\label{section:discussion}

In this article, we present a graphical formalism for representing gluonic operators and establish a systematic framework for constructing primitive evanescent operators, which is a crucial step in building the basis of full-color dressed gluon operators. By implementing this methodology, we efficiently generate primitive operators with lengths up to 7. This advancement allows us to systematically derive a complete gluon operator basis spanning operators up to dimension 14.

The construction of full-color dressed gluon operators requires two additional steps. The first step involves generating kinematic operators by adding D-pairs to the primitive operators. The second step entails dressing these kinematic operators with appropriate (single trace and multiple trace) color factors to form gauge-invariant operators. 
In the first step, the basis of length-3 kinematic operators can be conveniently established by adding all possible D-pairs to primitive operators. Nevertheless, for operators with a length exceeding three, such a construction leads to redundancy among the operators. This feature can be readily comprehended through the disparity between the gauge invariant basis and the primitive operators.
The gauge invariant basis can be constructed by the following gauge invariant building blocks \cite{Boels:2018nrr},
\begin{equation}
\begin{aligned}
&\bold{C}_{ij}=(p_i\cdot p_j)(\epsilon_i\cdot \epsilon_j)-(p_i\cdot \epsilon_j)(p_j\cdot \epsilon_i),\ \\
&\bold{A}_i^{i+1,a}=(p_i\cdot p_a)(p_{i+1}\cdot \epsilon_i)-(p_i\cdot p_{i+1})(p_a\cdot \epsilon_i),\ \\
\end{aligned}
\end{equation}
in which $a\ne i-1,i,i+1$.

For example, the length-5 gauge invariant basis contains 558 elements within the subsequent 3-classes,
\begin{equation}
\begin{aligned}
&\mathbf{C}_{12}\mathbf{C}_{34}\mathbf{A}_5^{1a_5}:45,\ \mathbf{C}_{12}\mathbf{A}_3^{4a_3}\mathbf{A}_4^{5a_4}\mathbf{A}_5^{1a_5}: 270,\ 
&\mathbf{A}_1^{2a_1}\mathbf{A}_2^{3a_2}\mathbf{A}_3^{4a_3}
\mathbf{A}_4^{5a_4}\mathbf{A}_5^{1a_5}: 243 \ .\\
\end{aligned}
\end{equation}
However, the lenght-5 primitive operator basis contains 1097 elements. Therefore, one can find at least 1097-558=539 linear relations among the primitive operators $P_a$, which will be called zero operators, and denoted by $Z_x$
\begin{equation}
Z_x=\sum_a f_{xa}P_a=0\ ,
\end{equation}
in which $f_{xa}$ are polynomial functions of $s_{ij}$.

The number of elements in the primitive operator basis and the gauge invariant basis, denoted by  $\mathcal{R}_n$ and $\mathcal{G}_n$ respectively, are given by

\medskip{}
\begin{tabular}{|c |c|c|c|c|c|c|}
\hline
$n$ & 2 &3&4&5&6&7  \\
\hline
$\mathcal{R}_n$ & 1 &4&54&1097&28705&955587  \\
\hline
$\mathcal{G}_n$ & 1 &4&43&558&8671&157400  \\
\hline
\end{tabular}.

\medskip{}
In order to  remove the redundancy, one may first construct all zero operators and subsequently eliminate the zero operators from the overcomplete operator basis through division, which can be faciliated with the aid of the Groebner basis.

Independent gauge-invariant operators can be systematically constructed by decomposing both kinematic factors and color factors into irreducible representations of the symmetric group $S_n$. Kinematic factors and color factors residing in the same irreducible representation are then combined to form gauge-invariant operators, ensuring independence through the symmetry properties of $S_n$.

A promising future direction involves systematically extending the framework of primitive operator methods to generic gauge theories (such as QED and QCD) and incorporating a broader class of operators, including fermionic, tensor, and non-local operators (e.g., Wilson loops and lines), both in the presence and absence of chiral symmetry breaking. Such an extension must address key challenges, such as establishing consistent diagrammatic representations for particles of different spins and developing systematic graphical rules to eliminate redundant operator diagrams.
It is also interesting to study whether the number of kinematic operators, or even the primitive operators, can be computed using the Hilbert series method \cite{Jenkins:2009dy, Hanany:2010vu, Lehman:2015via, Lehman:2015coa}.

\acknowledgments
We would like to thank Bo Feng for helpful discussions. 
The work of QJ is supported in part by the Science Challenge Project (No. TZ2025012), and NSAF No. U2330401. The work of GY is
supported in part by the National Natural Science Foundation of China (Grants No. 12425504,
12175291, 12047503, 12247103) and by the CAS under Grants No. YSBR-101. 
We also thank the support of the HPC Cluster of ITP-CAS.

\appendix

\section{Algebraic construction of primitive form factors}
\label{algebraic}

In this section, we analyze primitive operators from an algebraic perspective. Specifically, we establish a mapping between each kinematic form factor and a polynomial vector. Under this mapping, the set of kinematic form factors of fixed length forms a module. This framework enables an intuitive definition of primitive form factors: they correspond to the generators of the kinematic form factor module. Building on this definition, we provide a constructive method for obtaining primitive form factors. Although this method can be computationally less efficient than those detailed earlier, its underlying logic is straightforward. Moreover, it independently verifies primitive operators constructed via graphical rules in prior sections.

A general $n$-point kinematic form factor can be expressed as
\begin{align}
	\mathcal{F}^n=\sum_a z_af^n_a\,,\quad z_a\in \mathbb{R}[\{p_i\cdot p_j\}]\ ,
\end{align}
and $\mathbb{R}[\{p_i\cdot p_j\}]$ denotes the ring of real-coefficient polynomials in the variables $\{p_i\cdot p_j\}$. Here $\{f^n_a\}$ is the complete set of all multi-linear structures of polarization vectors, which we choose to be monomials. Each form factor is then naturally mapped to a polynomial vector in an $N$-dimensional space:
\begin{align}
	(z_1,z_2,\dots,z_N)\,,
\end{align} 
where $N$ is the number of multi-linear structures. We refer to this polynomial vector as a form factor vector.

Let us consider the case $n=2$ as an illustrative example. The initial step is to identify all $2$-point multi-linear structures $f^2_a$. These structures are given by:
\begin{align}
	f^2_a=\{p_1\cdot e_2\  p_2\cdot e_1, \ \ e_1\cdot e_2\}\,.
\end{align}
Consequently, each $2$-point form factor can be mapped to a polynomial 2-vector $(z_1,z_2)$. For example, the expression
\begin{align}
	(p_1\cdot e_2\ p_2\cdot e_1)-p_1\cdot p_2\ (e_2\cdot e_1)
\end{align}
corresponds to the vector $\left(1,-p_1\cdot p_2\right)$.

We say that a form factor vector $(z_1,\dots,z_N)$ is $\mathcal{F}$-homogeneous if all $z_a f^n_a$ are homogeneous polynomials of the same degree $k$. Here, the degree of a monomial is defined as the sum of the exponents of its variables $p_i$ and $e_i$ (e.g. $p_1\cdot e_2$ has degree 2). In this case, we call $(z_1,\dots,z_N)$ a degree-$k$ form factor vector. For example, in the length-2 case, $(1,-p_1\cdot p_2)$ is a dimension-4 form factor vector because both components are homogeneous polynomials of degree 4, where we also take into account of the degrees of multi-linear structures $f_a^n$'s. Such a vector corresponds precisely to the kinematic form factor of a dimension-$k$ gluon operator.

The set of length-$n$ form factor vectors is defined as the collection of polynomial $N$-vectors satisfying $n$ gauge invariance conditions:
\begin{align}
	\mathbb{F}^n=\Bigl\{(z_1,\dots,z_N)\in \mathbb{R}[\{p_i\cdot p_j\}]\ \Bigl|\ \sum_a z_a f^n_a\Big{|}_{e_k\to p_k}=0, k=1,2,\dots n\Bigr\}\,.
\label{Fn}
\end{align}
After setting $e_k \to p_k $, the expression $f_a^n$ becomes a multilinear structures of the remaining polarization vectors. Each gauge invariance condition then gives rise to a set of polynomial linear equations, corresponding to the coefficients of these multilinear structures:
\begin{align}
	\Bigl\{\sum_a c_{x,a} z_a=0\Bigr|x=1,\cdots, n_s\Bigr\}\,,\quad c_{x,a}\in \mathbb{R}[\{p_i\cdot p_j\}]\, ,
	\label{polyeqs}
\end{align}
here we use $n_s$ to denote the total number of such linear equations.

As an example, consider the length-2 case again. The form factor vector reads
\begin{align}
	\mathbb{F}^2=\Bigl\{(z_1,z_2)\in \mathbb{R}[\{p_1\cdot p_2\}]\ \Bigr|\ \sum_a z_a f^2_a\Big{|}_{e_i\to p_i}=0, i=1,2\Bigr\}\,.
	\label{originF2}
\end{align}
From the gauge invariance condition, we obtain the following equations:
\begin{align}
	p_1\cdot e_2(z_1\ p_1\cdot p_2+z_2)=0\,,\qquad
	p_2\cdot e_1(z_1\ p_1\cdot p_2+z_2)=0\,.
\end{align}
Thus \eqref{originF2} simplifies to
\begin{align}
	\mathbb{F}^2=\{(z_1,z_2)|z_i\in \mathbb{R}[s_{12}]\,,\ \{z_1\ p_1\cdot p_2+z_2=0\}\}\,,
\end{align}
since in this special case the two gauge conditions yield only one independent linear polynomial equation.

It is straightforward to see that for any $\mathcal{F}^n_1,\mathcal{F}^n_2\in \mathbb{F}^n$ and any $x_1,x_2\in \mathbb{R}[\{p_i\cdot p_j\}]$,
\begin{align}
	x_1\mathcal{F}^l_n+x_2\mathcal{F}^n_2 \in \mathbb{F}^n\,.
\end{align}
Thus $\mathbb{F}^n$ forms a module over the polynomial ring $\mathbb{R}[\{p_i\cdot p_j\}]$. In this framework, $\mathbb{F}^n$ is precisely the Syzygy module (see e.g. \cite{Cox:2015ode}) of the system \eqref{polyeqs}.  An essential property is that for any two equations in this system,
\begin{align}
	\text{deg}(c_{k_1,a})=\text{deg}(c_{k_2,a})\,,
\end{align}
since the substitution $e_i\to p_i$ preserves the degree of each $f^n_a$. Furthermore, each $c_{k,a}$ is a monomial. 
Due to these facts, $\mathbb{F}^n$ admits a generating set whose elements are homogeneous with respect to polynomial degree. This homogeneity arises naturally because all gluon operators can be classified by dimension, and the form factor of a dimension-$k$ operator is intrinsically a homogeneous polynomial of degree $k$.

So far we have illustrated the idea to understand the set of $n$-point form factors as a module defined as \eqref{Fn}. Along this line, one can define the primitive form factors as following:
\paragraph{Definition 1}\
The set of length-$n$ primitive form factors $\mathbb{P}^n=\{\mathcal{P}^n_i\}$ constitutes a generating set of $\mathbb{F}^n$, where each $\mathcal{P}^n_i$ is $\mathcal{F}$-homogeneous. Besides, the primitive form factors should be `linear independent', in the sense that no set of coefficients $\{c_i\in \mathbb{R}[\{s_{ij}\}]\}$ satisfies both the following conditions simultaneously:
\begin{itemize}
	\item[1] $\sum_i c_i \mathcal{P}^n_i= 0$.
	\item[2] At least one $c_i$=1,
\end{itemize}
which is equivalent to stating that $\mathcal{P}^n_i$ lies in the module generated by the other form factors. This definition aligns with prior characterizations of primitive form factors.

Based on this definition, we present a direct construction method for primitive form factors. First, we solve the syzygy module of the linear equations in \eqref{polyeqs} to obtain a generating set. Practically, this is achieved using the computer algebra system Singular \cite{DGPS}, which automatically produces an $\mathcal{F}$-homogeneous generating set:
\begin{align}
	\{\mathcal{P}^n_1,\dots,\mathcal{P}^n_A\}\,.
	\label{origingen}
\end{align}
But the generating set given by Singular may be overcomplete, so it is necessary to select a subset of \eqref{origingen} that satisfies Definition 1.

The elimination of linear relations proceeds recursively as follows. Suppose we have already examined some minimal-dimension elements within \eqref{origingen}
\begin{align}
	\{\mathcal{P}^l_{i_1},\dots,\mathcal{P}^l_{i_B}\}\,,
	\label{tested}
\end{align}
which exhibit no linear relations as specified in Definition 1. We compute the Grobner basis $\mathcal{G}$ of the module generated by \eqref{tested}. Next, we select an element $\mathcal{P}^n_k$ from the remaining candidates, prioritizing those of minimal dimension. Using polynomial division, we reduce $\mathcal{P}^n_k$ modulo $\mathcal{G}$.
\begin{enumerate}
\item If the remainder is zero, it implies a linear dependence:
\begin{align}
	\exists \{c_i\}\,,\ \mathcal{P}^n_k=\sum_j c_j \mathcal{P}^n_{i_j}\,.
	\label{lr}
\end{align}
In this case, $\mathcal{P}^n_k$is discarded, and the process continues with the next element.

\item If the remainder is non-zero, $\mathcal{P}^n_k$ is added to the set \eqref{tested} as $\mathcal{P}^n_{i_{m+1}}$.

\end{enumerate}
Crucially, this elimination proceeds from lower to higher dimensions because the relation \eqref{lr} requires that $\mathcal{P}^n_k$ (on the left) has strictly higher dimension than all $\mathcal{P}^n_{i_j}$ on the right. Consequently, an operator $\mathcal{P}^n_k$ can only be tested against elements of lower dimension.

By implementing this algebraic algorithm, we derived primitive form factors for lengths 2 through 5. The resulting count of these form factors coincides precisely with the number of corresponding primitive operators obtained via the operator diagram method in Section \ref{subsection:len<=7}, serving as a verification of the method's consistency.

\section{The Maschke's theorem}
\label{Maschke}
The decomposition of a representation space in presence of an invariant subspace can be achieve by using the Maschke's theorem.

Let $\rho$ be a representation of $G$, and $V$ is the representational space. Suppose $V$ has a non-trivial $G$-invariant subspace $V_1$, and decompose $V=V_1\oplus W$. Let $\mathcal{A}_0$ be the function from $V$ to $V_1$. For $v_1\in V_1,\ w\in W$, $\mathcal{A}_0 (v_1+w)=v_1$.

The operator is defined as
\begin{equation}\label{A-define}
\mathcal{A}=\frac{1}{|G|}\sum_{g\in G}\rho(g^{-1})\mathcal{A}_0\rho(g)
\end{equation}
Then $V$ can be decomposed into the direct sum of two $G$-invariant spaces:
\begin{equation}
V=\text{Im}\mathcal{A}\oplus \text{Ker}(\mathcal{A})
=V_1\oplus \text{Ker}(\mathcal{A})\ .
\end{equation}

\section{Algorithm to construct diagrams with $\mathcal{R}(T)\ge 1$}
\label{appendix:construct}
In this appendix, we present the algorithm for deriving the complete set of diagrams with $n$ vertices that satisfy $\mathcal{R}(T)\ge 1$. 
In Section \ref{subsection:diagram-with-prim}, we prove that the loop numbers for these diagrams satisfy $L\le n-1$. The only diagram attaining $L=n-1$ is depicted in Figure \ref{fig:operator-upper-bound-no-arrow}. 
In Section \ref{appendix:algorithm}, we discuss the algorithm for deriving diagrams with $L\le n-2$, which completes overall derivation.

\subsection{The upper bound of loop number}
\label{subsection:diagram-with-prim}

For a diagram $T$ with $n$-vertices to have at least one primitive configuration, ($\mathcal{R}(T)\ge 1$), its loop number $L$ cannot be excessively large. This constraint arises because higher loop numbers correlate with increased edge counts and a greater number of type-vv vertices, leading to heightened complexity in the skeleton diagram and a tendency for its loop number to rise. Crucially, $\mathcal{R}(T)=0$ holds for any connected skeleton diagram where loop number larger than one, as discussed in Section \ref{subsubsection:loop}. 

We find that $L$ has the following upper bound:

\textbf{In order for a diagram $T$ with $n$-vertices to have at least one primitive configuration, the loop number $L$ must satisfy}
\begin{equation}\label{loop-bound}
L-n\le -1\ .
\end{equation}

To begin, we prove \eqref{loop-bound} assuming both $T$ and its skeleton diagram $T_0$ are connected. We start with a connected skeleton diagram $T_0$, and try to construct an undircted diagram $T$ which has largest $L-n$. 
We use $n_0$ and $L_0$ and to denote the number of vertices and loops of $T_0$. 
If $n_0=0$, then $T$ contains no $\text{deg}\ge3$ vertices, and it must be a 1-loop diagram with $n\ge2$, therefore \eqref{loop-bound} holds. From now on, we assume $n_0\ge 1$. If $n_0=1$, then $T_0$ must be a tree diagram, therefore $L_0=0$. If $n_0\ge 2$, then $L_0\le 1$ so that $\mathcal{R}(T_0)\ge 1$. In both case $L_0-n_0\le -1$.

The digram $T$ is constructed from $T_0$ by pairing the external legs of $T_0$ and connecting each pair with a path of successively adjacent degree-2 vertices. Each such connection increases the loop number by one and adds at least one new vertices. Therefore, the number of new vertices must be at least the number of new loops, and the diagram satisfies
\begin{equation}
L-L_0\le n-n_0,
\end{equation}
from which we derive
\begin{equation}\label{l-n-1}
L-n\le  L_0-n_0\le -1\ .
\end{equation}
So \eqref{loop-bound} holds for connected diagrams with connected skeleton diagram.

Next we consider the case $T$ is connected, but the skeleton diagram has $T_0$ has $\chi_0>1$ components. We can still follow the proof of $\chi_0=1$ case, and it can be found that:
\begin{equation}
L-n\le 1-2\chi_0\ .
\end{equation}

If $T$ is not connected, and has $\chi$ components, then
\begin{equation}
L=\sum_{i=1}^{\chi}L_i\le \sum_{i=1}^{\chi}(n_i-1)=n-\chi.\ 
\end{equation}
So we have proved \eqref{loop-bound} for generic operator diagrams, with equality possible only when both $T$ and $T_0$ are connected.

Next, we discuss the diagrams satisfying $L=n+1$.
Using \eqref{l-n-1}, one finds that it is only possible if
\begin{equation}
L-n=L_0-n_0=-1\ .
\end{equation}
When $n_0=1$ (an example is illustrated in Figure \ref{fig:np-loop}(c)), the upper bound can not be saturated. In this case, the diagram consists of a single vertex $T_0$ with degree at least 4, connected to paths of successively adjacent degree-2 vertices. Each path must contain at least two vertices, otherwise it degenerates into a skeleton snail as depicted in \ref{fig:np-loop}(a). Applying the derivation method for \eqref{l-n-1}, we obtain the following relation:
\begin{equation}
 L_0-n_0\le  -2\ .
\end{equation}

\begin{figure}
\centering
\includegraphics[scale=0.8]{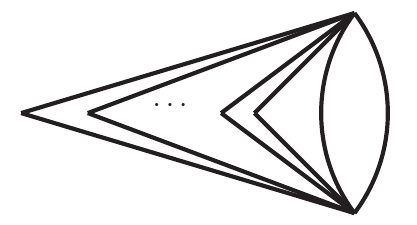}
\caption{Operator diagrams satisfying $L= n-1$.}
\label{fig:operator-upper-bound-no-arrow}
\end{figure}

Since $\mathcal{R}(T)\ge 1$ infers $L_0\le 1$, if $n_0\ge 3$, 
\begin{equation}
L_0-n_0\le 1-3=-2\ .
\end{equation}
Consequently, to reach the upper bound in \eqref{loop-bound}, the diagram must satisfy
\begin{equation}
n_0=2,L_0=1, L-L_0= n-n_0.
\end{equation}
The skeleton diagram must be a 1-loop bubble, and the undirected diagram must have the form of Figure \ref{fig:operator-upper-bound-no-arrow}.

\subsection{Diagrams with $L\le n-2$}
\label{appendix:algorithm}

\begin{figure}
\centering
\includegraphics[scale=0.6]{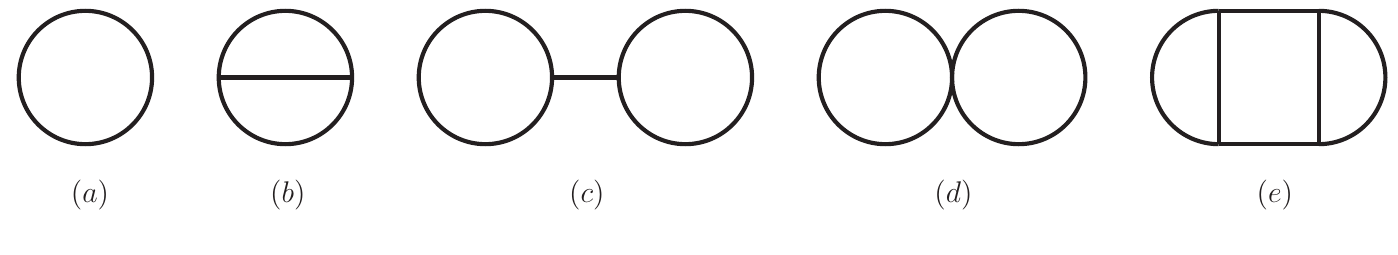}
\caption{Some vacuum topologies with $L\le 3$ and $\text{deg}(v_i)\ge 3$.}
\label{fig:len5vacuum}
\end{figure}

To complete the algorithm, we now consider diagrams satisfying $\mathcal{R}(T)\ge1$ and $L\le n-2$.
To begin, we construct the required skeleton diagrams. These diagrams contain no degree-2 vertices, except for those within skeleton snails. Upon removing all such degree-2 vertices from the skeleton snails, the skeleton diagrams can be regarded as subgraphs of vacuum diagrams with loop number $L\le n-2$ and no degree-2 vertices. Consequently, the first step is to generate connected vacuum diagrams $\{\mathcal{V}_i\}$ such that $L\le n-2$ and the vertex count $V\le n$, allowing only vertices with degree at least 3. For example, for $n=5$,  some diagrams with $L\le 3$ are illustrated in Figure \ref{fig:len5vacuum}.

\begin{figure}
\centering
\includegraphics[scale=0.5]{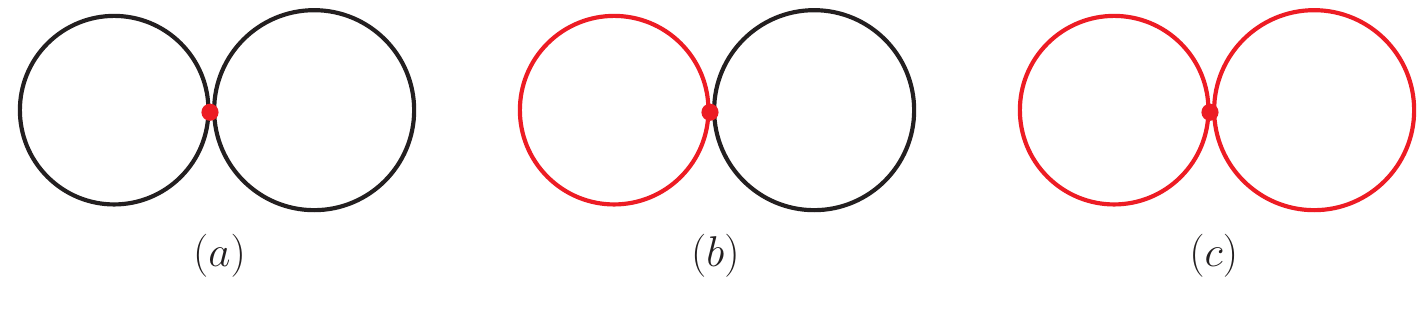}
\caption{Candidate stripped skeleton diagrams (colored red) of Figure \ref{fig:len5vacuum}(d).}
\label{fig:candidate-skeleton}
\end{figure}

Let $\mathbf{E}=\{e_1,\cdots, e_m\}$ be the edge set of $\mathcal{V}_i$. Each subset of  $S\subseteq\mathbf{E}$, combined with all vertices of $\mathcal{V}_i$, forms a candidate skeleton diagram. For example, the candidate skeleton diagrams for Figure \ref{fig:len5vacuum}(d) are presented in Figure \ref{fig:candidate-skeleton}, where we have omitted one diagram due to its symmetry with Figure \ref{fig:candidate-skeleton}(b).

If a self-loop structure (where an edge connects a vertex to itself) 
is included in the candidate skeleton diagram, one attaches a new vertex to each self-loop to construct a skeleton snail. Subsequently, all skeleton diagrams with $\mathcal{R}=0$ must be removed. For instance, Figure \ref{fig:candidate-skeleton}(c) exhibits $\mathcal{R}=0$ since it is a 2-loop skeleton diagram. New vertices are then introduced to edges outside the skeleton diagram: two new vertices per self-loop and one new vertex per non-self-loop edge.  Assuming the diagram now contains $n_1$ vertices, it represents an undirected diagram corresponding to length-$n_1$ operators. 

\begin{figure}
\centering
\includegraphics[scale=0.7]{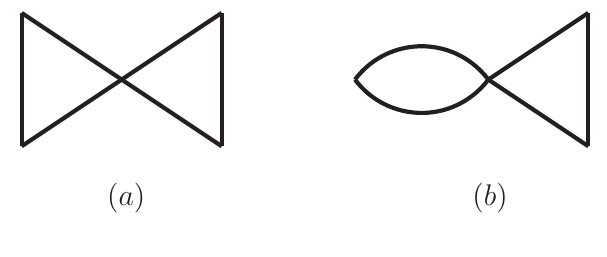}
\caption{Undirected diagrams obtained from Figure \ref{fig:candidate-skeleton}.}
\label{fig:gen-skeleton}
\end{figure}

The diagrams in Figure \ref{fig:gen-skeleton} are generated from Figure \ref{fig:candidate-skeleton}, as shown in the corresponding subfigures. Specifically, subfigure (a) of Figure \ref{fig:candidate-skeleton} generates subfigure (a) of Figure \ref{fig:gen-skeleton}, and subfigure (b) corresponds to subfigure (b), whereas subfigure (c) does not generate any diagram.

\begin{figure}
\centering
\includegraphics[scale=0.7]{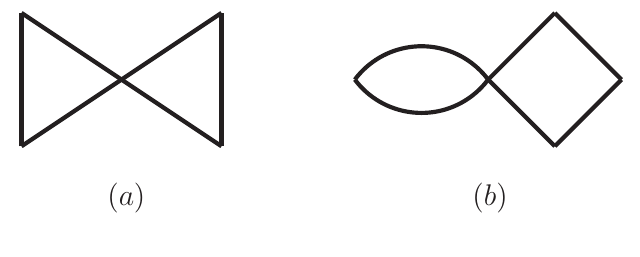}
\caption{Undirected diagrams after distributing $n-n_1$ vertices.}
\label{fig:length-5-final}
\end{figure}

Finally, the remaining $n-n_1$ vertices are distributed to the type-22 and  type-2v edges external to the skeleton diagram. For $n=4$, the resultant diagrams are provided in Figure \ref{fig:length-5-final}.

There are also disconnected operator diagrams when $n\ge4$. They can be constructed by combining several lower length connected operator diagrams.


\section{Further details for length-5 operators}
\label{appendix:len-5}
In this Appendix, we provide some details on length-5 primitive operators.
Length-5 operator diagrams with primitive configurations can be generated using the algorithm in Section \ref{subsection:diagram-with-prim}. The undirected diagrams are shown in Figure \ref{fig:len5op1}, and their symmetry groups are sumerized in Table \ref{symmetry-group-length-5}.

The primitive configurations of all undirected diagrams are given in Table \ref{len-5-prim}, in which 
\begin{equation}
\begin{aligned}
P_{33}^2=&F_{12}F_{34}F_{56}\Bigl[\{D_{13}F_{57},D_{67}F_{24}\}
+\frac{2}{3}\{D_{17}F_{35},D_{47}F_{26}\}\Bigr]\\
\end{aligned}
\end{equation}
$P_{33}^2$ generates a $\ydiagram{2,1}\otimes \ydiagram{2}$ representation of $S_3\times S_2$. It satisfies
\begin{equation}
(12)P_{33}^2=P_{33}^2,\ 
P_{33}^2+(123)P_{33}^2+(312)P_{33}^2=0,
\end{equation}

\begin{table}
\centering
\begin{tabular}{|c |c|c|c|c|}
\hline
$G_T$ & $T$& $\mathcal{R}(T)$& $\mathcal{R}(S_5T)$& Primitive configurations \\
\hline
$D_5$ & $T_{11}$&1&12&$F_{12}F_{23}F_{34}F_{45}F_{51}$\\
\hline
$S_1\times D_4$&$T_{21}$&3&45&\tabincell{c}{$P_{21}^1=F_{12}F_{23}F_{45}F_{56}D_{13}F_{46},$\\
$(13)(24)P_{21}^1,$\\ $\{F_{12},F_{23}\}\{F_{45},F_{56}\}D_{14}F_{36}$}  \\
\hline
$S_3\times S_2$&$T_{31}$&1&10&$F_{12}F_{23}F_{31}F_{45}F_{45}$  \\
\hline
$S_3\times S_2$&$T_{32}$&4&40&
\tabincell{c}{$P_{32}^1=F_{12}F_{34}F_{56}D_{1}F_{35}D_{2}F_{46},$\\$(12)P_{32}^1,\ (13)P_{32}^1,\ $
\\$F_{12}F_{34}F_{56}\{D_{1}F_{35},D_{4}F_{26}\}$}  \\
\hline
$S_3\times S_2$&$T_{33}$&5&50&
\tabincell{c}{$P_{33}^1=F_{12}F_{34}F_{56}D_{13}F_{57}D_{24}F_{67},$\\$(13)P_{33}^1,\ (23)P_{33}^1,P_{33}^2,(123)P_{33}^2$} \\
\hline
$S_3\times S_2$&$T_{34}$&1&10&$F_{12}F_{34}F_{56}D_{135}F_{78}D_{246}F_{78}$  \\
\hline
$S_1\times S_2\times S_2$&$T_{41}$&1 & 30&$F_{12}D_1F_{34}D_2F_{34}F_{56}F_{56}$\\
\hline
$S_1^3\times S_2$&$T_{51}$&1&60&
$F_{12}F_{34}D_{56}F_{34}F_{16}F_{25}$\\
\hline
$S_1^3\times S_2$&$T_{52}$&1&60&
$F_{12}D_3F_{12}D_{4}F_{35}\{F_{46},F_{56}\}$\\
\hline
$S_1^3\times S_2$&$T_{53}$&1&60&
$F_{12}D_1F_{34}D_{256}F_{34}F_{57}F_{67}$\\
\hline
$S_1^3\times S_2$&$T_{54}$&1&60&
$F_{12}D_{345}F_{12}D_{6}F_{47}\{F_{36},F_{57}\}$\\
\hline
$S_1^3\times S_2$&$T_{55}$&1&60&
$D_3F_{12}D_{45}F_{12}D_{6}F_{37}\{F_{46},F_{57}\}$\\
\hline
$S_2\times S_1$&$T_{61}$&1&60& 
$F_{13}F_{24}D_{3}F_{56}D_{4}F_{56}F_{12}$\\
\hline
$S_2\times S_1$&$T_{62}$&1&60&$F_{12}F_{34}D_{5}F_{12}D_{6}F_{34}F_{56}$\\
\hline
$S_2\times S_1$&$T_{63}$&3&180& 
\tabincell{c}{$F_{12}F_{23}D_{1}F_{46}D_{3}F_{56}F_{45},$\\$F_{12}F_{23}D_{4}F_{16}D_{5}F_{36}F_{45},$
\\$[1+(12)(34)]F_{12}F_{23}D_{1}F_{46}D_{5}F_{36}F_{45}$}\\
\hline
$S_2\times S_1$&$T_{64}$&1&60&$F_{12}F_{23}D_{14}F_{67}D_{35}F_{67}F_{45}$\\
\hline
$S_2\times S_1$&$T_{65}$&2&120& 
\tabincell{c}{$F_{12}F_{34}D_1F_{57}D_4F_{67}D_{23}F_{56},$\\$[1-(12)(34)]F_{12}F_{34}D_5F_{17}D_4F_{67}D_{23}F_{56}$}\\
\hline
$(S_1)^5$ &$T_{71}$&1&120& $F_{12}F_{34}D_2F_{45}D_1F_{67}D_{35}F_{67}$\\
\hline
\end{tabular}
\caption{Length-5 primitive operators.}
\label{len-5-prim}
\end{table}

In the table, $\mathcal{R}(S_5T)$ denotes the total number number of configurations under the action of vertex permutations, as discussed in Section \ref{subsection:len<=7}. For example, the symmetry group of $T_{21}$ is $S_1\times D_4$, which contains 8 elements. The set $S_5T_{21}$ includes $\frac{5!}{8}$ independent undirected diagrams, and each undirected diagram has $\mathcal{R}(T_{21})$ primitive configurations. Thus, the total number of primitive operators generated from $T_{21}$ is given by
\begin{equation}
\mathcal{R}(S_5T_{21})=\frac{5!}{8}\mathcal{R}(T_{21})=45\ .
\end{equation}

By summing over $\mathcal{R}(S_5T)$ values in Table \ref{len-5-prim}, we find that there are a total of 1097 primitive operators of length-5.


\bibliographystyle{JHEP}
\bibliography{/Users/jin/Documents/tex/PSUThesis/Biblio-Database}{}

\providecommand{\href}[2]{#2}\begingroup\raggedright\begin{thebibliography}{10}

\bibitem{Jin:2020pwh}
Q.~Jin, K.~Ren and G.~Yang, \emph{{Two-Loop anomalous dimensions of QCD
  operators up to dimension-sixteen and Higgs EFT amplitudes}},
  \href{https://doi.org/10.1007/JHEP04(2021)180}{\emph{JHEP} {\bfseries 04}
  (2021) 180} [\href{https://arxiv.org/abs/2011.02494}{{\ttfamily
  2011.02494}}].

\bibitem{Fermi:1934hr}
E.~Fermi, \emph{{An attempt of a theory of beta radiation. 1.}},
  \href{https://doi.org/10.1007/BF01351864}{\emph{Z. Phys.} {\bfseries 88}
  (1934) 161}.

\bibitem{Huggett:1995vya}
N.~Huggett and R.~Weingard, \emph{{The renormalisation group and effective
  field theories}}, \href{https://doi.org/10.1007/BF01063904}{\emph{Synthese}
  {\bfseries 102} (1995) 171}.

\bibitem{vanKolck:1999mw}
U.~van Kolck, \emph{{Effective field theory of nuclear forces}},
  \href{https://doi.org/10.1016/S0146-6410(99)00097-6}{\emph{Prog. Part. Nucl.
  Phys.} {\bfseries 43} (1999) 337}
  [\href{https://arxiv.org/abs/nucl-th/9902015}{{\ttfamily nucl-th/9902015}}].

\bibitem{Weinberg:1980wa}
S.~Weinberg, \emph{{Effective Gauge Theories}},
  \href{https://doi.org/10.1016/0370-2693(80)90660-7}{\emph{Phys. Lett. B}
  {\bfseries 91} (1980) 51}.

\bibitem{Gupta:2012mi}
R.S.~Gupta, H.~Rzehak and J.D.~Wells, \emph{{How well do we need to measure
  Higgs boson couplings?}},
  \href{https://doi.org/10.1103/PhysRevD.86.095001}{\emph{Phys. Rev. D}
  {\bfseries 86} (2012) 095001}
  [\href{https://arxiv.org/abs/1206.3560}{{\ttfamily 1206.3560}}].

\bibitem{Dawson:2013bba}
S.~Dawson et~al., \emph{{Working Group Report: Higgs Boson}},  in
  \emph{{Snowmass 2013}: {Snowmass on the Mississippi}}, 10, 2013
  [\href{https://arxiv.org/abs/1310.8361}{{\ttfamily 1310.8361}}].

\bibitem{LHCHiggsCrossSectionWorkingGroup:2016ypw}
{\scshape LHC Higgs Cross Section Working Group} collaboration, \emph{{Handbook
  of LHC Higgs Cross Sections: 4. Deciphering the Nature of the Higgs Sector}},
  \href{https://doi.org/10.23731/CYRM-2017-002}{\emph{CERN Yellow Rep. Monogr.}
  {\bfseries 2} (2017) 1} [\href{https://arxiv.org/abs/1610.07922}{{\ttfamily
  1610.07922}}].

\bibitem{D0:2012kms}
{\scshape D0} collaboration, \emph{{Measurement of the W Boson Mass with the D0
  Detector}}, \href{https://doi.org/10.1103/PhysRevLett.108.151804}{\emph{Phys.
  Rev. Lett.} {\bfseries 108} (2012) 151804}
  [\href{https://arxiv.org/abs/1203.0293}{{\ttfamily 1203.0293}}].

\bibitem{Boggia:2017hyq}
M.~Boggia et~al., \emph{{The HiggsTools handbook: a beginners guide to decoding
  the Higgs sector}}, \href{https://doi.org/10.1088/1361-6471/aab812}{\emph{J.
  Phys. G} {\bfseries 45} (2018) 065004}
  [\href{https://arxiv.org/abs/1711.09875}{{\ttfamily 1711.09875}}].

\bibitem{Weinberg:1979sa}
S.~Weinberg, \emph{{Baryon and Lepton Nonconserving Processes}},
  \href{https://doi.org/10.1103/PhysRevLett.43.1566}{\emph{Phys. Rev. Lett.}
  {\bfseries 43} (1979) 1566}.

\bibitem{Abbott:1980zj}
L.F.~Abbott and M.B.~Wise, \emph{{The Effective Hamiltonian for Nucleon
  Decay}}, \href{https://doi.org/10.1103/PhysRevD.22.2208}{\emph{Phys. Rev. D}
  {\bfseries 22} (1980) 2208}.

\bibitem{Buchmuller:1985jz}
W.~Buchmuller and D.~Wyler, \emph{{Effective Lagrangian Analysis of New
  Interactions and Flavor Conservation}},
  \href{https://doi.org/10.1016/0550-3213(86)90262-2}{\emph{Nucl. Phys. B}
  {\bfseries 268} (1986) 621}.

\bibitem{Grzadkowski:2010es}
B.~Grzadkowski, M.~Iskrzynski, M.~Misiak and J.~Rosiek, \emph{{Dimension-Six
  Terms in the Standard Model Lagrangian}},
  \href{https://doi.org/10.1007/JHEP10(2010)085}{\emph{JHEP} {\bfseries 10}
  (2010) 085} [\href{https://arxiv.org/abs/1008.4884}{{\ttfamily 1008.4884}}].

\bibitem{Lehman:2014jma}
L.~Lehman, \emph{{Extending the Standard Model Effective Field Theory with the
  Complete Set of Dimension-7 Operators}},
  \href{https://doi.org/10.1103/PhysRevD.90.125023}{\emph{Phys. Rev. D}
  {\bfseries 90} (2014) 125023}
  [\href{https://arxiv.org/abs/1410.4193}{{\ttfamily 1410.4193}}].

\bibitem{Liao:2016hru}
Y.~Liao and X.-D.~Ma, \emph{{Renormalization Group Evolution of Dimension-seven
  Baryon- and Lepton-number-violating Operators}},
  \href{https://doi.org/10.1007/JHEP11(2016)043}{\emph{JHEP} {\bfseries 11}
  (2016) 043} [\href{https://arxiv.org/abs/1607.07309}{{\ttfamily
  1607.07309}}].

\bibitem{Li:2020gnx}
H.-L.~Li, Z.~Ren, J.~Shu, M.-L.~Xiao, J.-H.~Yu and Y.-H.~Zheng, \emph{{Complete
  set of dimension-eight operators in the standard model effective field
  theory}}, \href{https://doi.org/10.1103/PhysRevD.104.015026}{\emph{Phys. Rev.
  D} {\bfseries 104} (2021) 015026}
  [\href{https://arxiv.org/abs/2005.00008}{{\ttfamily 2005.00008}}].

\bibitem{Murphy:2020rsh}
C.W.~Murphy, \emph{{Dimension-8 operators in the Standard Model Effective Field
  Theory}}, \href{https://doi.org/10.1007/JHEP10(2020)174}{\emph{JHEP}
  {\bfseries 10} (2020) 174}
  [\href{https://arxiv.org/abs/2005.00059}{{\ttfamily 2005.00059}}].

\bibitem{Li:2020xlh}
H.-L.~Li, Z.~Ren, M.-L.~Xiao, J.-H.~Yu and Y.-H.~Zheng, \emph{{Complete set of
  dimension-nine operators in the standard model effective field theory}},
  \href{https://doi.org/10.1103/PhysRevD.104.015025}{\emph{Phys. Rev. D}
  {\bfseries 104} (2021) 015025}
  [\href{https://arxiv.org/abs/2007.07899}{{\ttfamily 2007.07899}}].

\bibitem{Liao:2020jmn}
Y.~Liao and X.-D.~Ma, \emph{{An explicit construction of the dimension-9
  operator basis in the standard model effective field theory}},
  \href{https://doi.org/10.1007/JHEP11(2020)152}{\emph{JHEP} {\bfseries 11}
  (2020) 152} [\href{https://arxiv.org/abs/2007.08125}{{\ttfamily
  2007.08125}}].

\bibitem{Wilczek:1977zn}
F.~Wilczek, \emph{{Decays of Heavy Vector Mesons Into Higgs Particles}},
  \href{https://doi.org/10.1103/PhysRevLett.39.1304}{\emph{Phys. Rev. Lett.}
  {\bfseries 39} (1977) 1304}.

\bibitem{Shifman:1979eb}
M.A.~Shifman, A.I.~Vainshtein, M.B.~Voloshin and V.I.~Zakharov,
  \emph{{Low-Energy Theorems for Higgs Boson Couplings to Photons}},
  {\emph{Sov. J. Nucl. Phys.} {\bfseries 30} (1979) 711}.

\bibitem{Dawson:1990zj}
S.~Dawson, \emph{{Radiative corrections to Higgs boson production}},
  \href{https://doi.org/10.1016/0550-3213(91)90061-2}{\emph{Nucl. Phys. B}
  {\bfseries 359} (1991) 283}.

\bibitem{Djouadi:1991tka}
A.~Djouadi, M.~Spira and P.M.~Zerwas, \emph{{Production of Higgs bosons in
  proton colliders: QCD corrections}},
  \href{https://doi.org/10.1016/0370-2693(91)90375-Z}{\emph{Phys. Lett. B}
  {\bfseries 264} (1991) 440}.

\bibitem{Kniehl:1995tn}
B.A.~Kniehl and M.~Spira, \emph{{Low-energy theorems in Higgs physics}},
  \href{https://doi.org/10.1007/s002880050007}{\emph{Z. Phys. C} {\bfseries 69}
  (1995) 77} [\href{https://arxiv.org/abs/hep-ph/9505225}{{\ttfamily
  hep-ph/9505225}}].

\bibitem{Chetyrkin:1997sg}
K.G.~Chetyrkin, B.A.~Kniehl and M.~Steinhauser, \emph{{Strong coupling constant
  with flavor thresholds at four loops in the MS scheme}},
  \href{https://doi.org/10.1103/PhysRevLett.79.2184}{\emph{Phys. Rev. Lett.}
  {\bfseries 79} (1997) 2184}
  [\href{https://arxiv.org/abs/hep-ph/9706430}{{\ttfamily hep-ph/9706430}}].

\bibitem{Feruglio:1992wf}
F.~Feruglio, \emph{{The Chiral approach to the electroweak interactions}},
  \href{https://doi.org/10.1142/S0217751X93001946}{\emph{Int. J. Mod. Phys. A}
  {\bfseries 8} (1993) 4937}
  [\href{https://arxiv.org/abs/hep-ph/9301281}{{\ttfamily hep-ph/9301281}}].

\bibitem{Buras:1989xd}
A.J.~Buras and P.H.~Weisz, \emph{{QCD Nonleading Corrections to Weak Decays in
  Dimensional Regularization and 't Hooft-Veltman Schemes}},
  \href{https://doi.org/10.1016/0550-3213(90)90223-Z}{\emph{Nucl. Phys. B}
  {\bfseries 333} (1990) 66}.

\bibitem{Dugan:1990df}
M.J.~Dugan and B.~Grinstein, \emph{{On the vanishing of evanescent operators}},
  \href{https://doi.org/10.1016/0370-2693(91)90680-O}{\emph{Phys. Lett. B}
  {\bfseries 256} (1991) 239}.

\bibitem{Herrlich:1994kh}
S.~Herrlich and U.~Nierste, \emph{{Evanescent operators, scheme dependences and
  double insertions}},
  \href{https://doi.org/10.1016/0550-3213(95)00474-7}{\emph{Nucl. Phys. B}
  {\bfseries 455} (1995) 39}
  [\href{https://arxiv.org/abs/hep-ph/9412375}{{\ttfamily hep-ph/9412375}}].

\bibitem{Buras:1998raa}
A.J.~Buras, \emph{{Weak Hamiltonian, CP violation and rare decays}},  in
  \emph{{Les Houches Summer School in Theoretical Physics, Session 68: Probing
  the Standard Model of Particle Interactions}}, pp.~281--539, 6, 1998
  [\href{https://arxiv.org/abs/hep-ph/9806471}{{\ttfamily hep-ph/9806471}}].

\bibitem{Jin:2022ivc}
Q.~Jin, K.~Ren, G.~Yang and R.~Yu, \emph{{Gluonic evanescent operators:
  classification and one-loop renormalization}},
  \href{https://arxiv.org/abs/2202.08285}{{\ttfamily 2202.08285}}.

\bibitem{Jin:2022qjc}
Q.~Jin, K.~Ren, G.~Yang and R.~Yu, \emph{{Gluonic evanescent operators:
  two-loop anomalous dimensions}},
  \href{https://doi.org/10.1007/JHEP02(2023)039}{\emph{JHEP} {\bfseries 02}
  (2023) 039} [\href{https://arxiv.org/abs/2208.08976}{{\ttfamily
  2208.08976}}].

\bibitem{Jin:2023cce}
Q.~Jin, K.~Ren, G.~Yang and R.~Yu, \emph{{Is Yang-Mills theory unitary in
  fractional spacetime dimensions?}},
  \href{https://doi.org/10.1007/s11433-024-2370-6}{\emph{Sci. China Phys. Mech.
  Astron.} {\bfseries 67} (2024) 271011}
  [\href{https://arxiv.org/abs/2301.01786}{{\ttfamily 2301.01786}}].

\bibitem{Jin:2023fbz}
Q.~Jin, K.~Ren, G.~Yang and R.~Yu, \emph{{Gluonic evanescent operators:
  negative-norm states and complex anomalous dimensions}},
  \href{https://doi.org/10.1007/JHEP09(2024)151}{\emph{JHEP} {\bfseries 09}
  (2024) 151} [\href{https://arxiv.org/abs/2312.08445}{{\ttfamily
  2312.08445}}].

\bibitem{Boels:2018nrr}
R.H.~Boels, Q.~Jin and H.~Luo, \emph{{Efficient integrand reduction for
  particles with spin}},  \href{https://arxiv.org/abs/1802.06761}{{\ttfamily
  1802.06761}}.

\bibitem{Jenkins:2009dy}
E.E.~Jenkins and A.V.~Manohar, \emph{{Algebraic Structure of Lepton and Quark
  Flavor Invariants and CP Violation}},
  \href{https://doi.org/10.1088/1126-6708/2009/10/094}{\emph{JHEP} {\bfseries
  10} (2009) 094} [\href{https://arxiv.org/abs/0907.4763}{{\ttfamily
  0907.4763}}].

\bibitem{Hanany:2010vu}
A.~Hanany, E.E.~Jenkins, A.V.~Manohar and G.~Torri, \emph{{Hilbert Series for
  Flavor Invariants of the Standard Model}},
  \href{https://doi.org/10.1007/JHEP03(2011)096}{\emph{JHEP} {\bfseries 03}
  (2011) 096} [\href{https://arxiv.org/abs/1010.3161}{{\ttfamily 1010.3161}}].

\bibitem{Lehman:2015via}
L.~Lehman and A.~Martin, \emph{{Hilbert Series for Constructing Lagrangians:
  expanding the phenomenologist's toolbox}},
  \href{https://doi.org/10.1103/PhysRevD.91.105014}{\emph{Phys. Rev. D}
  {\bfseries 91} (2015) 105014}
  [\href{https://arxiv.org/abs/1503.07537}{{\ttfamily 1503.07537}}].

\bibitem{Lehman:2015coa}
L.~Lehman and A.~Martin, \emph{{Low-derivative operators of the Standard Model
  effective field theory via Hilbert series methods}},
  \href{https://doi.org/10.1007/JHEP02(2016)081}{\emph{JHEP} {\bfseries 02}
  (2016) 081} [\href{https://arxiv.org/abs/1510.00372}{{\ttfamily
  1510.00372}}].

\bibitem{Cox:2015ode}
D.A.~Cox, J.~Little and D.~O'Shea, \emph{{Ideals, Varieties, and Algorithms. An
  Introduction to Computational Algebraic Geometry and Commutative Algebra}},
  Undergraduate Texts in Mathematics, Springer (2025),
  \href{https://doi.org/10.1007/978-3-031-91841-4}{10.1007/978-3-031-91841-4}.

\bibitem{DGPS}
W.~Decker, G.-M.~Greuel, G.~Pfister and H.~Sch\"onemann, ``{\sc Singular}
  {4-4-0} --- {A} computer algebra system for polynomial computations.''
  \url{http://www.singular.uni-kl.de}, 2024.

\end{thebibliography}\endgroup

\end{document}